\documentclass{aa}  

\usepackage{graphicx}
\usepackage{txfonts}
\usepackage{lipsum}
\usepackage{subcaption}         
\usepackage{lscape}             
\usepackage{placeins}           
                                
\usepackage{txfonts}
\newcommand{\be}{\begin{eqnarray}}
\newcommand{\en}{\end{eqnarray}}

\usepackage[dvipsnames]{xcolor}
\usepackage{natbib}
\usepackage{url}
\usepackage{enumitem}
\usepackage{bm}

\usepackage{float}
\usepackage[colorlinks=true, linkcolor={blue!70!black}, citecolor={blue!70!black}, urlcolor={blue!70!black}]{hyperref}

\usepackage{threeparttablex} 
\usepackage{tabularx}        
\usepackage{longtable}     
\usepackage{comment}

\newcommand{\eps}{\varepsilon}
\newcommand{\bnabla}{\bm{\nabla}}
\newcommand{\bcdot}{\bm{\cdot}}
\newcommand{\rmn}{\mathrm}

\newcommand{\ergs}{\rm{ erg \,s}^{-1}}
   
\newcommand{\cc}{\rm~{cm}^{-3}}
\newcommand{\kpc}{\rm~kpc}

\usepackage{xcolor}

\usepackage{xpatch}
\usepackage{hyperref}
\hypersetup{
	colorlinks=true,
	breaklinks=true,
	citecolor=blue,
	allcolors=blue,
	frenchlinks=true
}

\makeatletter
\xpatchcmd\NAT@citex
{%
	\@citea\NAT@hyper@{%
		\NAT@nmfmt{\NAT@nm}%
		\hyper@natlinkbreak{\NAT@aysep\NAT@spacechar}{\@citeb\@extra@b@citeb}%
		\NAT@date
	}%
}
{%
	\@citea
	\NAT@nmfmt{\NAT@nm}%
	\NAT@aysep\NAT@spacechar
	\NAT@hyper@{\NAT@date}%
}
{}{}
\xpatchcmd\NAT@citex
{%
	\@citea\NAT@hyper@{%
		\NAT@nmfmt{\NAT@nm}%
		\hyper@natlinkbreak{\NAT@spacechar\NAT@@open\if*#1*\else#1\NAT@spacechar\fi}%
		{\@citeb\@extra@b@citeb}%
		\NAT@date
	}%
}
{
	\@citea
	\NAT@nmfmt{\NAT@nm}%
	\NAT@spacechar\NAT@@open\if*#1*\else#1\NAT@spacechar\fi
	\NAT@hyper@{\NAT@date}%
}
{}{}
\makeatother

\makeatletter
\renewcommand*\aa@pageof{, page \thepage{} of \pageref*{LastPage}}
\makeatother

\begin{document} 

\title{Identifying heating processes in simulations with an entropy-based scheme: A single jet episode in a galaxy cluster}
\titlerunning{Identifying heating processes in galaxy clusters}

  \author{M.\ Meenakshi \inst{\ref{instAIP}}
        \and 
        R. Weinberger\inst{\ref{instAIP}}
        \and
        C. Pfrommer\inst{\ref{instAIP}}
        \and
        T. Berlok\inst{\ref{instNBI}} 
        }
    \institute{
    Leibniz Institute for Astrophysics, An der Sternwarte 16, D-14482 Potsdam, Germany \label{instAIP}        \\ \email{mmeenakshi@aip.de}
    \and 
    Niels Bohr Institute, University of Copenhagen, Blegdamsvej 17, 2100 Copenhagen, Denmark \label{instNBI}
        }

   \date{\today}

  \abstract{
    Understanding heating processes in galaxy clusters is essential for predicting the regulation of radiative cooling and star formation, and for clarifying the mechanisms underlying active galactic nucleus (AGN) feedback in cool-core clusters. We investigate the processes through which AGN jets deposit heat into the intracluster medium (ICM) by tracking passive entropy scalars in magneto-hydrodynamic (MHD) simulations. This enables us to systematically disentangle the contributions from different heating channels. We successfully validate this method with several idealized tests, including turbulent heating, heating by anisotropic Braginskii viscosity, dissipative and adiabatic heating by shocks using in-situ shock-detection methods, and cosmic ray (CR) heating through the excitations and damping of Alfv\'en waves. Using this methodology, we simulate single-epoch outbursts of high-power jets with varying densities in a cluster environment. Light jets produce wider bubbles, displacing a larger fraction of the gas in the cluster's core, whereas comparatively denser jets propagate more efficiently to larger distances without significantly disturbing the central region. During early evolution, shock heating of the ICM prevails irrespective of jet density. At later times, light jets primarily heat the ICM through turbulently induced heating (which includes dissipation of vortices emerging from a turbulent cascade and mixing gas of different temperatures) and compressible dissipation of sound waves and very weak shocks, while denser jets continue to dissipate most of their energy via shocks. Particularly, turbulently induced heating by solenoidal-shear motions prevails inside the cocoon, whereas shocks and acoustic compressions dominate outside. In light jets, the forward shock weakens rapidly, whereas dense jets can sustain strong bow shocks to large distances. This heating estimator allows us to identify the dominant heating mechanism responsible for resolving the cooling flow problem in future self-regulated AGN jet simulations.}

  \keywords{Galaxies: active - Galaxies: jets - Galaxies: clusters: Intracluster medium - Methods: numerical}

   \maketitle

   \nolinenumbers


\section{Introduction}
A substantial fraction of galaxy clusters host cool cores, characterized by high central electron densities ($n_\mathrm{e}\gtrsim 10^{-2}\, {\rm cm^{-3}}$), short cooling times ($\lesssim$1 Gyr), low central temperatures $k_\mathrm{B} T_\mathrm{e}$ (with the central value typically a factor of 3 lower than the maximum value at intermediate cluster radii) and low central cluster entropies ($K_\mathrm{e}=k_\mathrm{B} T_\mathrm{e} n_\mathrm{e}^{-2/3}<30\,\mathrm{keV}\,\mathrm{cm}^2$, see \citealt{hudson_2010}). Under these conditions, strong cooling flows are expected to develop \citep{fabian_1994}, fueling vigorous star formation in the brightest cluster galaxies. However, such large-scale gas inflows and associated star formation are not observed \citep{peterson_2003,sanders_2008,gitti_2012} -- a long-standing discrepancy commonly referred to as the cooling flow problem.

To explain this discrepancy, theorists have proposed the presence of a heating mechanism that regulates radiative cooling in cluster cores \citep[see, for a recent review,][]{bourne_2023}. Feedback from active galactic nuclei (AGN), particularly in the form of relativistic jets, is widely considered the most plausible mechanism, supported by both observations and numerical simulations \citep[][and many more]{rafferty_2006,sullivan_2011,gaspari_2012a, gaspari_2012b,fornasiero_2025}. Observations of radio lobes or X-ray cavities seen in several cool-cores \citep[see e.g.][]{david_2009, blanton_2011,gendron_2021} indicate that the jets can break out of the host's interstellar medium and propagate to large distances. These jets carry relativistic particles, inflate hot bubbles, and generate shocks within the cluster core. Although the energy output of AGN feedback is sufficient to offset radiative cooling, it cannot transform cool-core clusters into non-cool-core (NCC) systems on the buoyancy timescale because the mechanical energy couples only weakly to the intracluster medium (ICM) \citep{pfrommer_2012}. Nevertheless, AGN feedback may still play an indirect role in the transformation of cool-core clusters into NCC systems by contributing to the cluster merger process \citep{chen_2026}.

A number of simulations have attempted to understand the evolution of large-scale jets across different densities, powers, and episodes of jets in various cluster environments \citep[e.g.,][]{perucho_2014,weinberger_2017,ehlert_2018, english_2019,perucho_2023,husko_2023,tsai_2025,stewart_2025}. Nonetheless, how exactly the jet's energy couples to the ICM and results in heating remains uncertain. Several mechanisms have been proposed that can contribute to this heating, including mixing of hot bubble gas with the ICM \citep{hillel_2016,hillel_2017}, turbulence dissipation \citep{Kunz_2011,zhuravleva_2014,zhuravleva_2016}, weak shocks or sound waves \citep{fabian_2003,ruszkwoski_2004,yang_2016a,tang_2017}, thermal conduction \citep{yang_2016b}, and cosmic-ray (CR) heating \citep{pfrommer_2013,ruszkwoski_2017,ruszkowski&pfrommer_2023}. However, the complexity of jet--ICM interactions demands a detailed and physically consistent framework to separate the various heating channels. 

Earlier simulation studies have explored this problem using a variety of complementary approaches. \citet{yang_2016a} and \citet{hillel_2016} used tracer particles to follow the evolution of particle energies and to characterize heating, cooling, and energy transport processes. \citet{li_2017} estimated the conversion of kinetic energy into thermal energy using an artificial viscosity scheme, separating the contributions from shocks and turbulent dissipation. \citet{martizzi_2019} applied an entropy tracer-based analysis to quantify jet-induced heating in hydrodynamic simulations. While these studies provide valuable insights, they have not included several key physical processes, most notably magnetic fields and CRs, that are essential for a complete picture of jet--ICM coupling. For instance, \citet{ruszkwoski_2017} showed that CR transport and streaming can be crucial for efficient ICM heating by AGN jets. Taken together, these findings underscore the need for both sophisticated physical modeling and robust numerical diagnostics to assess jet-driven heating accurately.

In this work, we implement an entropy-based heating estimator within the moving mesh code \textsc{Arepo} \citep{springel_2010,pakmor_2016}, which adopts the original idea of an entropy tracer-based heating algorithm in \textsc{Athena} \citep{ressler_2015,martizzi_2019}. This work focuses on describing and validating this implementation, identifying and studying its numerical limitations, and subsequently applying it to investigate heating processes driven by jets of constant power but varying density contrasts in galaxy clusters. Additionally, to examine shock heating induced by jets, we utilize the already implemented shock finder \citep{schaal_2015} and analyze the corresponding heating estimates. Studies have shown that jets with different densities evolve distinctively, leading to variations in their interaction with the ICM \citep{guo_2015,ehlert_2023} and, consequently, in their heating mechanisms. We therefore examine a range of jet-to-ambient density ratios to explore how these differences influence the jet evolution and the spatial distribution of heating. A detailed analysis of self-regulated jets will be presented in a forthcoming paper.

The structure of the paper is as follows. In Sect.~\ref{sec:theory}, we describe the theoretical framework and discuss the implementation of our heating estimation algorithm in Sect.~\ref{sec:algorithm}. The validation tests for this method are presented in Sect.~\ref{sec:tests}, with a particular focus on turbulent heating by numerical viscosity (Sect.~\ref{sec:turb_heating}) and Braginskii-viscosity (Sect.~\ref{sec:brag_heating}), CR heating in Sect.~\ref{sec:cr_heating}, with tests of the shock finder and the associated shock heating (Sect.~\ref{sec:shock_heating_test}), and reversible adiabatic heating (Sec~\ref{sec:adiabatic_heating_test}). We then examine the heating produced by single-jet outbursts in dense cluster environments in Sect.~\ref{sec:jet_test}, and summarize our conclusions in Sect.~\ref{sec:conclusions}.

\section{Theory}
\label{sec:theory}

In this section, we introduce the various heating processes and discuss how we can infer their contributions from numerical solutions of the hyperbolic differential equations governing magneto-hydrodynamics (MHD). This includes turbulent heating (either numerically or via anisotropic Braginskii viscosity), shock heating (dissipative and adiabatic), CR heating, and adiabatic heating as a result of compressible motions. 

\subsection{Thermal energy and entropy equations}
\label{sec:theory_equations}
We start with the thermal energy equation,
\be
\frac{\partial \eps}{\partial t} + \bnabla \bcdot (\eps \bm{\varv}) = - P \bnabla \bcdot\bm{\varv} + \mathcal{H} - \mathcal{C}, 
\label{eq:int_ene_eqn}
\en
where the thermal energy density ($\varepsilon$) and pressure ($P$) are related via the equation of state, $\eps=P/(\gamma-1)$, where $\gamma=5/3$ is the adiabatic index. The mean gas velocity is $\bm{\varv}$, and the cooling rate density is $\mathcal{C}$, and the heating rate density is given by
\be
\mathcal{H} = \mathcal{H}_\mathrm{tu} + \mathcal{H}_\mathrm{cr},
\label{eq:heating_rate_dist}
\en
where $\mathcal{H}_\mathrm{tu}$ and $\mathcal{H}_\mathrm{cr}$ denote the heating rate density due to viscous dissipation of turbulent motions and due to collisionless damping of Alfv\'en waves that were excited by streaming CRs, respectively, which is detailed below.

Using the first law of thermodynamics, the equation for the specific entropy, $s$, is given as:
\be
\rho T \frac{{\rm d}s}{{\rm d}t} =\mathcal{H} - \mathcal{C}
\label{eq:heat_eqn}
\en
where ${\rm d}/{\rm d}t = \partial/\partial t + \bm{\varv} \bcdot \bnabla$ is the Lagrangian derivative. For convenience, we introduce the entropic function, which remains constant along an adiabatic curve:
\be
K = P\,\rho^{-\gamma}.
\en
The specific entropy $s$ is related to the pressure and gas mass density, $\rho$, via
\be
    s = c_V \log \left(\frac{K}{K_0}\right)
    = c_V \log \left(\frac{P_{\phantom{0}}\rho^{-\gamma}}{P_0^{}\rho_0^{-\gamma}}\right),
    \label{eq:specific_entropy}
\en
where $c_V = k_\mathrm{B}/[(\gamma-1) \mu m_\mathrm{p}]$ is the specific heat at constant volume, $k_\mathrm{B}$ is the Boltzmann constant, $m_\mathrm{p}$ is the proton mass, $\mu$ is the mean molecular weight, and the subscripts 0 denote the reference value for adiabatic changes. Eliminating the expression for $s$ in Eqs.~\eqref{eq:heat_eqn} and \eqref{eq:specific_entropy}, we obtain the net heating rate density in terms of pressure and density:
\be
 \mathcal{H} - \mathcal{C} =  \frac{P}{\gamma - 1} \frac{{\rm d} \log\left(P\rho^{-\gamma}\right)}{{\rm d}t} = \frac{\rho^\gamma}{\gamma - 1} \frac{{\rm d} \left(P\rho^{-\gamma}\right)}{{\rm d}t}.
 \label{eq:heat_rate_eqn}
\en

\subsection{Physical and numerical turbulent heating}
\label{sec:theory_turb_heating}
Viscous dissipation of subsonic turbulent energy occurs either because eddies at the grid scale are not numerically resolved or through the explicit modeling of physical viscosity. In the framework of ideal MHD, there is no explicit viscosity, so that dissipation of eddies is exclusively done via numerical viscosity. Associated with this energy dissipation is an increase in entropy directly analogous to physical entropy generation, giving rise to the turbulent heating rate density due to numerical viscosity, $\mathcal{H}_\mathrm{tu,nu}$. 

By contrast, if we explicitly model viscosity and ensure that the grid scale is everywhere below the dissipation scale, we speak about a direct numerical simulation. Depending on the properties of the fluid to be modeled, we can either use the Navier--Stokes equation with the coefficients for bulk and shear viscosity or -- in the case of weakly collisional media -- Braginskii MHD \citep{braginskii_1965}, yielding in both cases the viscous heating rate $\mathcal{H}_\mathrm{tu,ph}$ in Eq.~\eqref{eq:heating_rate_dist}. Hence, $\mathcal{H}_\mathrm{tu,nu}$ and $\mathcal{H}_\mathrm{tu,ph}$ describe the same physical effect of viscous dissipation and only differ in the numerical realization via implicit (numerical) or explicit (physical) viscosity, respectively. In a direct numerical simulation where the viscous scale is fully resolved everywhere, the physical viscosity should dominate even though it is impossible to completely eliminate discretization effects such as entropy mixing as a result of the cell averaging procedure in Godunov schemes, so that numerical dissipation should always be present at some level. In practice, the grid spacing in our simulations spans several orders of magnitude, making it impossible to resolve the viscous scale in large (cosmological) runs everywhere and therefore necessitating at least partial reliance on numerical dissipation.

In Braginskii MHD, ion momentum and electron heat are transported anisotropically along magnetic field lines, leading to anisotropic viscous dissipation and thermal conduction, respectively. Heat conduction is a spatial transport process but can also drive and damp motions via the magneto-thermal instability and thermal dissipation \citep{fabian_2005,zweibel_2018,Perrone_2022a,Perrone_2022b}. Here, we solely focus on viscous heating as a source of thermal energy. The corresponding heating rate involves the anisotropic viscosity tensor $\bm\Pi$ and reads as follows \citep{berlok_2020}:\footnote{Here the notation $\bm{:}$ denotes a double contraction, such that $\bm{b}\bm{b}\bm{:}\bnabla \bm{\varv} = \sum_i \sum_j b_i b_j \partial_i \varv_j$, which is equivalent to the trace of the matrix product of $\bm{b}\bm{b}$ and $\bnabla \bm{\varv}$.}
\begin{align}
\mathcal{H}_{\rm tu,ph} = -\,\bm{\Pi : \bnabla \varv} &= \frac{\rho \nu_\parallel}{3} 
\left( 3 \, \bm{b}\bm{b} \bm{:} \bnabla \bm{\varv} - \bnabla \bcdot \bm{\varv} \right)^2>0,
\end{align}
where $\nu_\parallel$ is the viscosity coefficient along the magnetic field, $\bm{b} = \bm{B}/B$ is the magnetic unit vector, $B$ is the magnitude of $\bm{B}$, and $\bm{b b}$ denotes the dyadic product of two vectors.

\subsection{Cosmic ray heating}
\label{sec:theory_cr_heating}
CRs introduce an additional relativistic fluid component that interacts with the thermal plasma via pressure gradients. CRs are advected with the gas and interact with plasma waves so that they diffuse and stream along the magnetic field relative to the plasma \citep{thomas_2019,ruszkowski&pfrommer_2023}. The evolution of the CR energy density, $\eps_{\rm cr}$, in the one-moment approximation is given by \citep{pfrommer_2017}
\begin{align}
\frac{\partial \eps_{\rm cr}}{\partial t} 
+ \bm{\bnabla} \bcdot 
\left[ \eps_{\rm cr} \bm{\varv} 
+ (\eps_{\rm cr}+P_{\rm cr}) \bm{\varv}_{\rm st}
-\kappa \bm{b}(\bm{b}\bcdot\bnabla \eps_{\rm cr})\right]
= \nonumber\\
- P_{\rm cr}\bm{\bnabla}\bcdot \bm{\varv}
+\bm{\varv}_\mathrm{st} \bcdot \bm{\bnabla} P_{\rm cr}
+\Lambda_{\rm cr},
\end{align}

where $P_{\rm cr}$ is the CR pressure, $\bm{\varv}_{\rm st}$ is the CR streaming velocity, and $\kappa$ is the CR diffusion coefficient along the magnetic field. The three source terms on the right-hand side denote adiabatic CR gains and losses, CR losses as a result of CR-driven plasma instabilities, and non-adiabatic gain and loss terms (denoted by $\Lambda_{\rm cr}$) as a result of CR acceleration at shocks as well as Coulomb and hadronic losses of CR energy, some of which will also be transferred to the thermal plasma. Because the Coulomb and hadronic heating rates are negligible in galaxy clusters on AGN jet duty cycles \citep{ruszkowski&pfrommer_2023}, we will not further discuss this here.\footnote{Note that this heating rate can be easily included in the heating rate analysis in case this becomes important, e.g., in galaxies.} 

Interactions between CRs and Alfv\'en waves lead to near-isotropization of CRs in the wave frame, causing them to stream along the magnetic field relative to the plasma with a velocity of
\begin{eqnarray}
  \label{eq:vstream}
  \bm{\varv}_{\rm st} = -\bm{\varv}_{\rm A}\, \mathrm{sgn}(\bm{B}\bcdot\bnabla P_{\rm cr})
  = -\frac{\bm{B}}{\sqrt{\rho}}\,
  \frac{\bm{B}\bcdot\bnabla P_{\rm cr}}{\left|\bm{B}\bcdot\bnabla P_{\rm cr}\right|},
\end{eqnarray}
where we adopt the Heaviside--Lorentz system of units. As streaming CRs excite Alfv\'en waves \citep{Kulsrud_1969,Shalaby_2021,Shalaby_2023}, which saturate via particle trapping \citep{Lemmerz_2025} and collisionless damping processes, they transfer energy to the thermal plasma at a rate \citep{pfrommer_2017}:
\be
\label{eq:heat_cr_eqn}
   \mathcal{H}_{\rm cr} = -\bm{\varv}_{\rm st} 
   \bcdot \bnabla P_{\rm cr} = \frac{1}{\sqrt{\rho}}\frac{(\bm{B} \bcdot \bnabla P_{\rm cr})^2}{|\bm{B} \bcdot \bnabla P_{\rm cr}|} > 0.
\en

\subsection{Dissipative and adiabatic shock heating}
\label{sec:theory_shock_heating}

Shocks increase the thermal energy and entropy of the gas through two processes:
\begin{enumerate}
\item \textit{Irreversible dissipative heating}, which arises from the entropy jump across the shock, and
\item \textit{Shock adiabatic heating}, resulting from the (in principle) reversible compression of the gas.
\end{enumerate}
The former represents heat added to the system due to entropy generation, while the latter corresponds to the volume work done on the gas by shock compression. However, the heating rate associated with either effect is not provided by an explicit source term in the energy equation (Eq.~\ref{eq:int_ene_eqn}), but instead arises from an integral constraint on the thermal energy \citep{landau_1987}. Thus, estimating the energy deposition at a shock requires a dedicated in-situ shock finder that communicates the pre- and post-shock thermal energies to the shock surface and calculates the corresponding heating rates \citep{schaal_2015}.

\begin{figure}
\centering
\includegraphics[width=\linewidth, keepaspectratio]{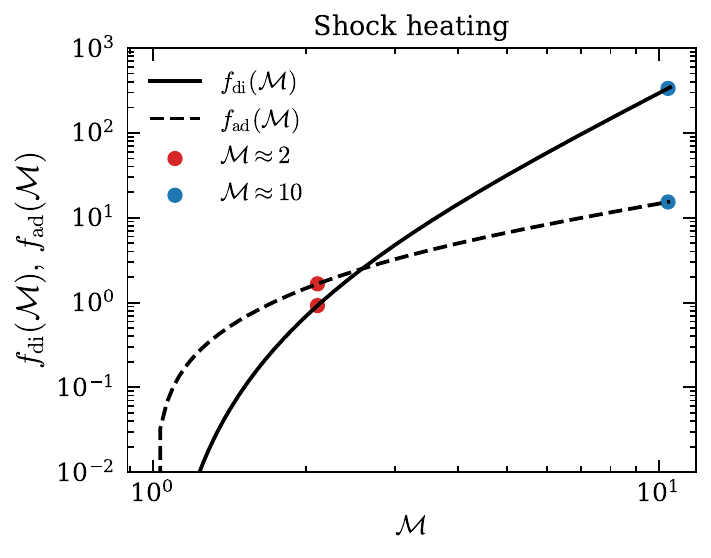}
\caption{Dissipative (Eq.~\eqref{eq:fdi}) and adiabatic heating (Eq.~\eqref{eq:fad})) rates at shocks as functions of the Mach number. The rates are normalized by the upstream thermal energy density times the sound speed to make them dimensionless. The markers indicate the heating corresponding to the weak ($\mathcal{M}\approx 2$) and strong shocks ($\mathcal{M}\approx 10$) in our tests.}
\label{fig:diss_ad_plot}
\end{figure}

Consider a shock with Mach number $\mathcal{M} = \varv_1/c_1$ propagating into a medium of density $\rho_1$, thermal energy density $\eps_1$, and sound speed $c_1$. The associated increase in thermal energy of the gas per unit area ($A$) and per unit time ($t$) resulting from shock dissipation and adiabatic compression, respectively, is given as \citep{pfrommer_2006,springel_2010}
\begin{align}
\label{eq:shock_diss}
    \frac{{\rm d}E_{\rm di}}{{\rm d}t {\rm d}A}  &= \frac{\rho_1 \varv_1\rho_2^{\gamma-1}}{\gamma -1}\left(K_2 - K_1 \right) ~~~= \eps_1 c_1 f_{\rm di} (\mathcal{M}), \\
\label{eq:shock_ad}
    \frac{{\rm d}E_{\rm ad}}{{\rm d}t {\rm d}A} &= \frac{\rho_1 \varv_1 K_1}{\gamma -1}\left(\rho_2^{\gamma-1} - \rho_1^{\gamma-1}  \right)  = \eps_1 c_1 f_{\rm ad} (\mathcal{M}),
\end{align}
where 
\begin{align}
& f_{\rm di}(\mathcal{M}) = \mathcal{M} \left[f_K(\mathcal{M}) - 1\right] f_{\rho}^{\gamma -1} (\mathcal{M}), 
\label{eq:fdi}\\
& f_{\rm ad} (\mathcal{M}) = \mathcal{M} \left[f_{\rho}^{\gamma -1}(\mathcal{M}) - 1\right]. 
\label{eq:fad}
\end{align}
Here, the subscripts 1 and 2 label pre- and post-shock quantities, respectively, that are defined in the shock rest frame. The jump in entropy and density across the shock is represented by the standard Rankine-Hugoniot jump conditions:
\begin{align}
f_\rho(\mathcal{M})&\equiv\frac{\rho_2}{\rho_1}
=\frac{(\gamma+1)\mathcal{M}^2}{(\gamma-1)\mathcal{M}^2+2},\\
f_K(\mathcal{M})&\equiv\frac{K_2}{K_1}
=\frac{2\gamma\mathcal{M}^2-(\gamma-1)}{\gamma+1}
\left[\frac{(\gamma-1)\mathcal{M}^2+2}{(\gamma+1)\mathcal{M}^2}\right]^{\gamma}.
\label{eq:RH_jump}
\end{align}
The corresponding dissipative and adiabatic heating rates, described by Eqs.~\eqref{eq:fdi} and \eqref{eq:fad}, are presented in Fig.~\ref{fig:diss_ad_plot}. The dissipative term dominates at high Mach numbers, while the adiabatic contribution becomes more significant for weaker shocks (i.e., low Mach numbers). The values corresponding to a weak and strong shock, with Mach numbers 2 and 10, respectively, are also indicated in the same figure. We have performed tests for shock detection and estimated the associated heating rates later in Sect.~\ref{sec:shock_heating_test}. For completeness, the heating rate densities at shocks are defined as:
\begin{equation}
\mathcal{H}_\mathrm{sh,di} = A\frac{{\rm d}\eps_{\rm di}}{{\rm d}t {\rm d}A}, \qquad\mbox{and}\qquad
\mathcal{H}_\mathrm{sh,ad} = A\frac{{\rm d}\eps_{\rm ad}}{{\rm d}t {\rm d}A},
\label{eq:H_sh}
\end{equation}
where ${\rm d}\eps_{\rm di}$ and ${\rm d}\eps_{\rm ad}$ correspond to an increase in the thermal energy density of the gas at the shock due to shock dissipation and adiabatic compression, respectively.

\subsection{Adiabatic heating/cooling}
Reversible heating arises from adiabatic processes, such as compression or expansion, that modify the thermal energy or temperature without generating net entropy. In our study, we distinguish heating from the continuous compressible motions and adiabatic heating at shocks (see Sect.~\ref{sec:theory_shock_heating}). We estimate the thermal energy change associated with such a continuous adiabatic process as
$$\mathcal{H}_{\rm ad} = -P \bnabla \bcdot \bm{\varv}, $$
which is positive upon compression over the region of interest. Since this quantity can be either positive or negative, we refer to this as the adiabatic heating/cooling rate density.

\section{Algorithm}
\label{sec:algorithm}

The conservative update of \textsc{Arepo} ensures that the total energy is conserved to machine precision. Even in the absence of physical viscosity or external energy input, artificial heating (entropy generation) arises from volume-averaging of vector quantities such as velocity and magnetic field. These errors manifest as numerical viscosity (and, in ideal MHD, numerical resistivity). Therefore, comparing a case in which entropy is merely advected (entropy conservation) to one where heating is included (energy conservation) provides a measure of the heat added to the system. 

Our implementation for estimating the heating rate builds on, and extends, the approach utilized in \textsc{Athena} \citep{ressler_2015,martizzi_2019}. Our approach employs two entropy-based passive scalars, $K=P \rho^{-\gamma}$ and $S=\log(K)$, with the former corresponding to the form used in those studies\footnote{Our passive scalar $S$ and the specific entropy $s$ are related via $S=s/c_V+\log K_0$, see Eq.~\eqref{eq:specific_entropy}.}. In this work, these entropy-based estimators are used exclusively to quantify turbulently-induced and weakly compressible dissipation (which includes heating via turbulent cascading, mixing, and viscous dissipation of compressible sound waves and very weak shocks). We find that both methods perform well for smooth problems; however, they diverge at contact discontinuities or sharp gradients, yielding negative heating rates due to numerical discretization errors. Therefore, we only consider the heating contribution in regions where both (formally equivalent) heating estimators indicate a positive value, and discard the corresponding cells otherwise (see details below and in Appendix~\ref{app:discretization_issue}).

For shocks, we utilize the shock-finder and shock-heating estimator implemented in \textsc{Arepo} \citep{schaal_2015}. The corresponding dissipative and adiabatic heating are calculated using the standard shock-jump relations (see Sect.~\ref{sec:theory_shock_heating}). Shock zones and surfaces, identified by the shock finder, are explicitly excluded from the turbulent-heating (and possible mixing heating) calculation. This separation ensures that shock dissipation and turbulent dissipation are treated independently without double-counting.

\subsection{Entropy-based heating estimator}
In theory, the heat added to the system can be measured either from the energy equation or from the entropy equation. The energy equation accounts for cell-based energy changes arising from Riemann fluxes, turbulent dissipation introduced by numerical viscosity, adiabatic expansion, compression, and any additional physical heating terms. However, disentangling these contributions is non-trivial. Therefore, we turn to the entropy formulation, which provides a more direct way to isolate the irreversible heating.

We define heating as the generation of entropy through the dissipation of kinetic and magnetic energy into thermal energy, either physically or numerically. The key idea to quantify this heating is to evaluate the entropy of the gas using two independent approaches:

\begin{enumerate}
    \item \textit{From energy conservation:} entropy is computed directly from the density and pressure obtained from the conservative hydrodynamic equations, and
  \item \textit{From entropy conservation:} entropy is evolved using a purely advection equation in the absence of explicit heating or cooling.
\end{enumerate}

The difference between these two entropy estimates quantifies the entropy generated in the system by irreversible processes and is used for the estimation of the heating rate due to numerical processes associated with turbulently induced and weakly compressible dissipation (which includes heating via turbulent cascading, mixing, and viscous dissipation of compressible sound waves and very weak shocks). Below, we describe the implementation of the entropy-based heating estimator.

\subsubsection{Numerical approach}
\label{sec:numerical_approach}

We use Eq.~\eqref{eq:heat_rate_eqn} to estimate the heating within the system, and the following section outlines its numerical implementation.

\paragraph{Passive scalars for entropy conservation}

We introduce two passive scalars for entropy as 

\begin{equation}
\begin{aligned}
    K &= P\,\rho^{-\gamma}, \quad {\rm and}\\
    s &= c_V \log \left(\frac{K}{K_0}\right),
\end{aligned}\label{eq:passive_scalars}
\end{equation}
where the last equation is the specific entropy (Eq.~\ref{eq:specific_entropy}).\footnote{In the implementation, we evolve $S=\log K$, as a proxy for the passive entropy scalar $s$.} These passive scalars are constant for an ideal adiabatic flow with no external heating or cooling (i.e. $\mathcal{H} = \mathcal{C} = 0$).  In Lagrangian form, this implies (from Eq.~\ref{eq:heat_rate_eqn})

\be
    \frac{{\rm d}s}{{\rm d}t} =  0 = \frac{{\rm d} K}{{\rm d} t} ,
    \label{eq:entro_conserve}
\en

and the corresponding Eulerian (conservative) forms are

\begin{equation}
\begin{aligned}
    \frac{\partial (\rho K)}{\partial t} + \bnabla \bcdot \left(\rho K \bm{\varv}\right) = 0 ,  \\
    \frac{\partial (\rho s)}{\partial t} + \bnabla \bcdot \left(\rho s \bm{\varv}\right) = 0 .
    \label{eq:cont_eq}
\end{aligned}
\end{equation}

The second term in the above equations represents the divergence of the flux of the scalar quantity transported by the flow and is solved alongside the Euler equations using a finite-volume approach. Note that we use the extrapolated values of $K$ at the cell interface when estimating the flux term for $s$.\\

\paragraph{Heating rate estimation}

Let $t^{(n)}$ and $t^{(n+1)}$ denote two successive time steps. At $t^{(n+1)}$, the conservative update of \textsc{Arepo} provides $\rho^{(n+1)}$ and $P^{(n+1)}$, from which we compute
\begin{equation}
\begin{aligned}
K_{\rm egy}^{(n+1)} &= P^{(n+1)}\left[\rho^{(n+1)}\right]^{-\gamma}
\quad\mbox{and}\\
s_{\rm egy}^{(n+1)} &= c_V \log\!\left(\frac{K_{\rm egy}^{(n+1)}}{K_0}\right).
\end{aligned}
\label{eq:new_entropy}
\end{equation}
These values include the effects of heating produced during the time step, including numerical dissipation. In parallel, the conservative forms of Eq.~\eqref{eq:cont_eq} are evolved, representing the entropy expected in the absence of cooling or dissipation ($K_{\rm adv}, s_{\rm adv}$). Thus, the change in the passive scalars over the time step is
\begin{equation}
\begin{aligned}
\Delta K &= {K_{\rm egy}}^{(n+1)} - {K_{\rm adv}}^{(n+1)},\\
\Delta s &= {s_{\rm egy}}^{(n+1)} - {s_{\rm adv}}^{(n+1)}.
\end{aligned}
\end{equation}
In the absence of radiative cooling, the heating rate density can be estimated as
\begin{equation}
\mathcal{H} = 
\begin{cases}
  \dfrac{(\rho_{\rm avg})^\gamma}{\gamma -1} \,\dfrac{\Delta K}{\Delta t}, \\[1em]
 \dfrac{\mu m_\mathrm{p}P_{\rm avg}}{k_\mathrm{B} }\,\dfrac{\Delta s}{\Delta t},
\end{cases}
\quad \Delta t = t^{(n+1)} - t^{(n)}
\label{eq:heating_rate}
\end{equation}
where $\rho_{\rm avg}$ and $P_{\rm avg}$ denote the average density and pressure computed from the previous\footnote{When refinement occurs, these values are copied from the parent cell to both child cells.} and current time-step values.

\subsubsection{Discretization effects on heating estimation}
\label{sec:discret_error}

 We find that the two heating-estimation methods ($K$ and $s$-based) show systematic differences due to discretization issues in finite-volume methods. To further investigate the occurrence of these discrepancies, we examine the numerical conditions that give rise to unphysical (locally negative or diverging) heating. Appendix~\ref{app:discretization_issue} shows that differences between the two methods stem from advection effects and sharp gradients. The advection errors occur particularly near contact discontinuities where low- and high-entropy material meet across sharp density gradients. In Sect.~\ref{sec:post_contact}, we demonstrate that, over a single time step, the advection of a contact discontinuity in $K$ and $s$ across our discretized mesh leads to diverging heating rate estimates (Fig.~\ref{fig:egy_post_contact}), producing negative heating in the former and positive heating in the latter. Thus, we confirm that these estimates are of numerical origin for steep density and hence, entropy discontinuities.

Additionally, in Sect.~\ref{sec:jet_first_order} we show that several of these non-physical heating estimates can be reduced by limiting gradients or adopting more diffusive reconstructions, effectively reverting the scheme to first-order (Fig.~\ref{fig:heating_K_S}). However, since these sharp contacts inevitably introduce some level of discrepancy, our analysis focuses on regions where both methods yield positive heating rates, which reliably indicate genuine physical heating. We adopt the $K$-based estimator as our primary diagnostic (as also used in \citealt{martizzi_2019}) for our analysis on heating by jets and use the $s$-based estimates as an additional check to verify that a given region corresponds to physically correct heating.

\subsection{Shock detection and heating estimation}
\label{sec:shock_theory}

We use the shock-finder implemented in \textsc{Arepo} by \citet{schaal_2015} to identify shocks and explicitly estimate the associated heating, as described in Sect.~\ref{sec:theory_shock_heating}. 

The dissipative and adiabatic fluxes are multiplied by the shock-surface area to estimate the rate at which heat is deposited into the gas at the shock. To estimate the shock surface, we adopt a spherical cell approximation in our study, where the area for the cell with volume $V$ is given as $(3V\sqrt{\pi}/4)^{2/3}$. We adopt a minimum Mach number threshold of $\mathcal{M}=1.3$ for identifying shock zones, following \citet{schaal_2015}, to avoid including numerical noise or false positives in compressive turbulent regions.

\subsection{Adiabatic heating/cooling}
\label{sec:ad_heating}
To compute the adiabatic heating/cooling due to continuous flows within a computational cell, we use the velocity divergence ($\bnabla \bcdot \bm{\varv}$) provided by \textsc{Arepo} outside of (numerically broadened) shocks. In the finite-volume formulation, this quantity is computed by summing the velocity projected along the outward normal vector over all cell faces ($f$), weighted by the corresponding face areas, and dividing by the cell volume ($V_i$). The corresponding mathematical expression is

$$
(\bnabla \bcdot \bm{\varv})_i
   = \frac{1}{V_i}
     \sum_{f}
     \left( \bm{\varv}_f \bcdot \bm{n}_f \right) A_f, \nonumber
$$
where $A_f$ is the face area and $\bm{n}_f$ is the face normal vector. In the case of shocks, we explicitly compute the shock adiabatic heating due to compression (see Eq.~\ref{eq:shock_ad}), as described in Sect.~\ref{sec:theory_shock_heating}.

\begin{figure*}
\centering
\includegraphics[width=\linewidth, keepaspectratio]{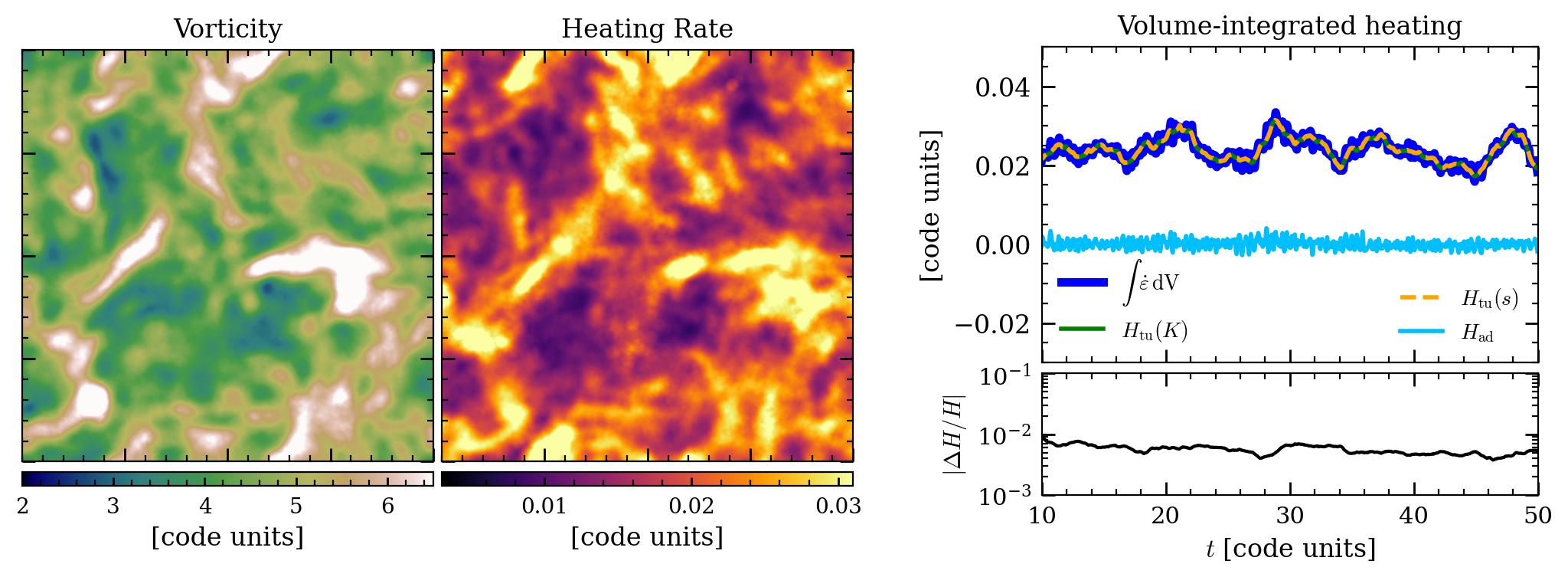}
\caption{\textit{Left}: Volume-weighted projected vorticity ($|\bnabla \times \bm{\varv}|$) and turbulent heating rate due to numerical viscosity, both weighted by volume, for the entire box width. \textit{Right}: Volume-integrated turbulent heating rates, $H_\mathrm{tu}(K)$ and $H_\mathrm{tu}(s)$, thermal energy gain rate, and the adiabatic term $H_\mathrm{ad}$ over the entire simulation box are shown in the top panel, while the bottom panel presents the relative error between the two heating-estimation methods, $\Delta H/H=H_\mathrm{tu}(K)-H_\mathrm{tu}(s)/H_\mathrm{tu}(K)$.}
 \label{fig:vorticity_heat}
\end{figure*}

\subsection{Algorithmic procedure for heating estimation}
In the previous sections, we described the numerical framework used to estimate heating from the various physical mechanisms. When applying this framework to full simulations (e.g., AGN jets in galaxy clusters), radiative cooling is also included and may, in principle, interfere with the heating estimate. In this case, the heating rate can still be estimated using the procedure described above, as operator splitting allows a clean separation of processes. Within each time-step, the following sequence is executed:
\begin{enumerate}[label=(\roman*)]
\item We execute the hydrodynamics solver along with the shock-finder. In cells with a negative velocity divergence that are labeled as shock zones by the shock finder, we estimate the heating (dissipative and adiabatic) using the formalism described in Sect.~\ref{sec:shock_theory}.
\item  We then compute the heating rate in the remaining cells (i.e., in all non-shock zone cells) with the entropy-based estimation method (Eq.~\ref{eq:heating_rate}) and estimate the continuous (i.e., non-shock) adiabatic heating term in those cells (see Sect.~\ref{sec:ad_heating}).
\item Next, we apply radiative cooling.
\item The passive scalars $K$ and $s$ are then updated, ensuring that the heating estimates do not include the effects of cooling.
\item Finally, we apply jet injection and any other energy input, e.g., from non-ideal heating sources such as CR--Alfv\'en wave heating and Braginskii viscosity heating at this stage.
\item The procedure is repeated from step (i).
\end{enumerate}

Volume integration of the heating rate density of each process $i$ over the domain of interest yields 
$$H_i = \int \mathcal{H}_i \, dV.$$

\section{Testing different heating mechanisms}
\label{sec:tests}

 In this section, we present several tests to validate the accuracy of the entropy-based heating estimation method and the existing shock-finder. The different heating mechanisms examined here are particularly relevant for quantifying the various heating processes of AGN jets to ICM heating, namely, turbulent and Braginskii viscosity heating, CR heating, and heating by low- and high-Mach-number shocks, and adiabatic heating due to continuous compressive motions.
 
For the entropy-based heating estimator, we compute heating rates $\mathcal{H}(K)$ and $\mathcal{H}(s)$ using both scalar quantities, $K$ and $s$, respectively. We compare the reliability of the two approaches in different tests here. While our algorithm typically employs these entropy-based heating estimators to quantify numerical turbulent heating and entropy mixing at the cell level, here we instead use them to assess the accuracy of heating-rate estimates for non-ideal processes in the absence of exact analytical solutions, such as Braginskii viscosity and CR heating.

\subsection{Turbulent heating by numerical viscosity}
\label{sec:turb_heating}
  
We investigate heating from forced turbulent driving in a three-dimensional (3D) periodic box with a resolution of $64^3$ cells. The parameters for the subsonic turbulence are adopted from \citet{bauer_2012}, corresponding to a Mach number $\mathcal{M}\sim0.3$ in Table~1 of their paper. The domain is a unit cube filled with an ideal gas ($\gamma=5/3$) initialized at uniform density ($\rho=1$) and pressure ($P=1$) in code units. We employ purely solenoidal driving, with the acceleration field generated in Fourier space over the range $k_{\rm min} = 6.27$ to $k_{\rm max} = 12.57$. The relative amplitudes of the forcing modes in this range follow a scaling of the power spectrum of $E(k) \propto k^{-5/3}$. The driving routine applies the external acceleration in two half-steps -- at the start and end of each timestep -- resulting in a leapfrog-type time-integration scheme.

A quasi-steady turbulent state is reached by $t\approx 10$, with a 3D velocity dispersion of ${\sim} 0.4$. We extend the run for 16 additional eddy-turnover times, with the turnover time defined at the injection scale. This corresponds to $\approx$~40 code units (one turnover time is ${\approx}~ 2.5$ code units). The corresponding plots for this test are shown in Fig.~\ref{fig:vorticity_heat}. The left panels display the volume-weighted projected vorticity ($|\bnabla \times \bm{\varv}|$) and heating rates, while the top right panel shows the evolution of the box-integrated heating rates overlaid over the volume-integrated increase in thermal energy (first term in Eq.~\ref{eq:int_ene_eqn}), the adiabatic term. The bottom right panel shows the relative error between the $K$ and $s$-based heating estimation methods. The close correspondence between vorticity and heating rate confirms that turbulence predominantly drives dissipation through vortical cascades.

In the right panels, the heating rates computed using both methods are in good agreement and consistent with the thermal energy change rate. For solenoidal forcing, the adiabatic term integrated over the entire domain is expected to be negligible, which is indeed confirmed in the simulation (blue curve). We find that as turbulence develops and begins to dissipate, the turbulence driving rate is partitioned between changes in kinetic energy and the heating rate (i.e., the thermal energy change in this case, as the adiabatic term is negligible). This energy balance is accurately recovered in our simulations, together with the corresponding heating rates.

\begin{figure}
\centering
\includegraphics[width=\linewidth, keepaspectratio]{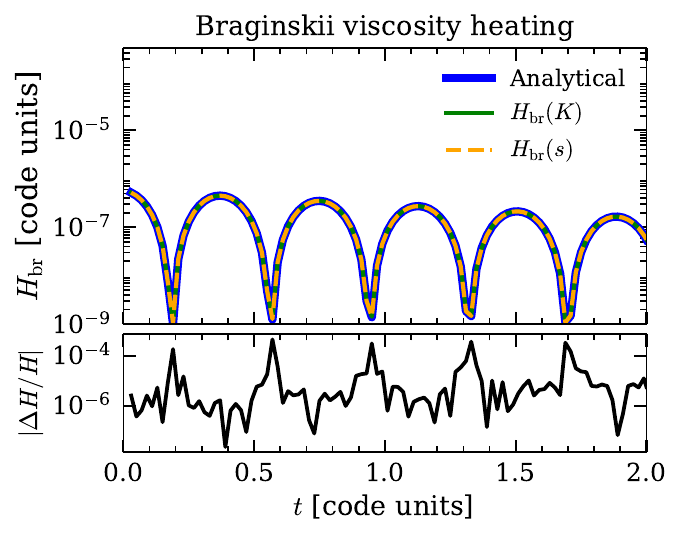}
\caption{Volume-integrated Braginskii viscosity heating as a function of time, $t$. The relative error between the two heating-estimation methods is shown in the bottom panel.}
\label{fig:brag_heat_2d}
\end{figure}

\subsection{Heating by Braginskii viscosity}
\label{sec:brag_heating}
Here, we closely follow the test from \citet[Sect. 4.3]{berlok_2020} and set up a standing fast magnetosonic wave in two-dimensional (2D) space for a box of length one in code units, and a resolution of $256^2$ grid points. The background magnetic field is set perpendicular to the plane, i.e., $\bm{B}= B_0\hat{z}$, where $B_0=1$, which corresponds to a plasma beta value of $\beta=P/(B^2/8\pi)=25$. The initial
density, magnetic field, and velocity are set as
\begin{align}
&\frac{\delta \rho}{\rho_{0}} = \frac{\delta B_{z}}{B_{0}}
= A \cos(\bm{k}\bcdot\bm{r}) \, \sin(\omega_{0} t)\, \rmn{e}^{-\gamma_{\rm br} t}, \\
&\bm{\varv}(\bm{r}, t) = -A \sin(\bm{k}\bcdot\bm{r}) 
\left[ \omega_{0}\cos(\omega_{0} t) - \gamma_{\rm br} \sin(\omega_{0} t) \right] 
\rmn{e}^{-\gamma_{\rm br} t/2} \frac{\bm k}{k^2},
\end{align}
where $\bm{k} = k_x \hat{\bm{e}}_x + k_y \hat{\bm{e}}_y$ is the wave number, with $k_x \equiv k_\perp = 2\pi/L$, $k_y = 0$, $L$ is the box length, and $A = 10^{-3}$ is the amplitude of perturbations. Here, $\omega_{0} = \mathrm{Re}(\omega)$ and $\gamma_{\rm br} = -\mathrm{Im}(\omega)$ 
are the real and imaginary parts of the complex frequency $\omega$, 
which is given by
\begin{equation}
\omega = \pm k_{\perp}\sqrt{\varv_{A}^{2} + {c_s}^{2} - \, {\left(\frac{k_{\perp} \nu_{\parallel}}{6}\right)}^2}
        - {\rm i} \, \frac{k_{\perp}^{2}\,\nu_{\parallel}}{6}, 
\end{equation}
where $\varv_{A}$ is the magnitude of Alfv\'en velocity, $c_s=\sqrt{\gamma P/\rho}$ is the adiabatic sound speed, and the Braginskii viscosity coefficient $\nu_{\parallel}$ is set to 0.05 here. 

The above equation indicates that Braginskii viscosity both damps at a rate $\gamma_{\rm br}$ as well as makes the waves dispersive. The viscosity leads to heating because the anisotropic viscous stress tensor does irreversible work against velocity gradients along the magnetic fields, converting bulk kinetic energy into thermal energy. The heating rate per unit volume here is given as\footnote{\rm We do not use the notation $\mathcal{H}_{\rm tu, ph}$ introduced in Sect.~\ref{sec:theory_turb_heating}, because, in this test, we simulate dissipation of a magnetosonic wave mediated by Braginskii viscosity rather than dissipation of turbulent motions.} (see Sec~\ref{sec:theory_turb_heating} for details),
\be \label{eq:brag_heat}
\mathcal{H}_{\mathrm{br}} = \frac{\rho \nu_{\parallel}}{3} 
\left(\bnabla \bcdot \bm{\varv} \right)^2.
\en
The heating rates derived from both entropy tracers, together with the analytical solution, are shown in Fig.~\ref{fig:brag_heat_2d}. The agreement between the entropy-based estimates and the theoretical prediction is excellent. 

The viscous heating exhibits oscillations because of the sinusoidal structure of the velocity field of the wave.  The zero-points in heating occur at times when the velocity amplitude is zero for the wave, i.e., 
$ \tan(\omega_{0} t)  = \omega_{0}/\gamma_{\rm br}$. In this test, $\omega_0 = 8.3$ and $\gamma_{\rm  br} = 0.329$, yielding zero-crossing times at approximately $t\approx 0.2, 0.56, 0.942, \ldots$, which is consistent with Fig.~\ref{fig:brag_heat_2d}. The amplitudes gradually decay due to the damping term.

\begin{figure}
\centering
\includegraphics[width=\linewidth, keepaspectratio]{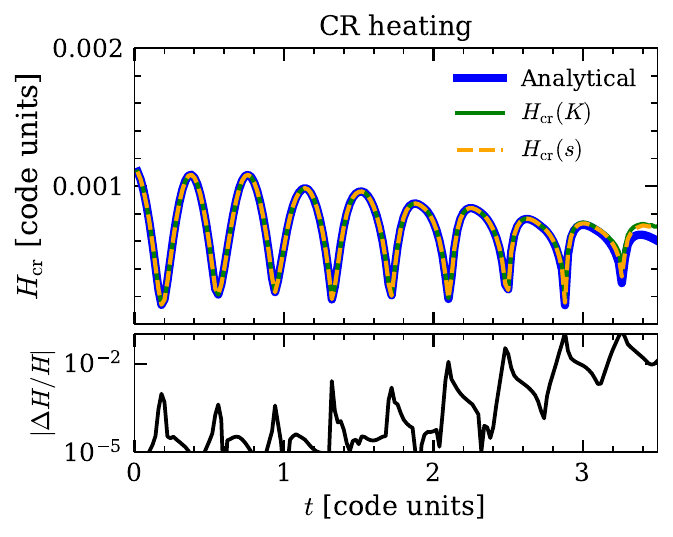}
\caption{Volume-integrated CR Alfv\'en-wave heating in a periodic 2D domain as a function of time, $t$. The relative error between the two heating-estimation methods is shown in the bottom panel.}
\label{fig:cr_heat_2d_3.5}
\end{figure}

\begin{figure*}
\centering
\includegraphics[width=\linewidth, keepaspectratio]{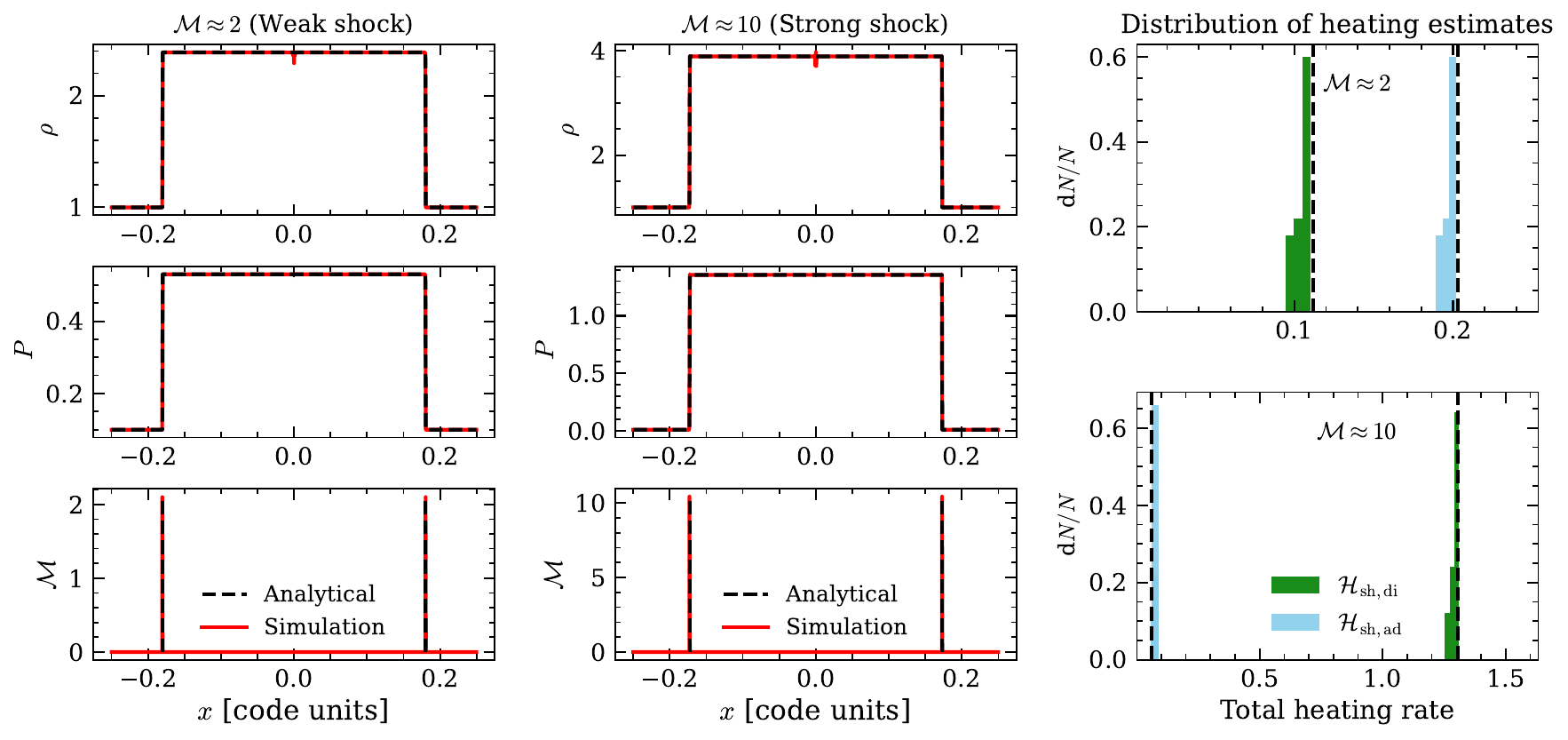}
\caption{\textit{Left and center}: Density (top), pressure (middle), and Mach number (bottom) profiles from a 1D shock test for weak and strong shocks. The black dashed curves show the analytical solution, while the red solid curves show the simulation results. \textit{Right}: We compare the distribution of the dissipative (green) and adiabatic (blue) shock heating (obtained from the two shock surfaces) against the analytical solution (black dashed).}
\label{fig:shock_test}
\end{figure*}

\subsection{Alfv\'en-wave-mediated CR heating}
\label{sec:cr_heating}
We set up a 2D periodic box, with $256^2$ grid points, to quantify the heating due to damping of Alfv\'en waves that are generated through a CR streaming instability \citep{Kulsrud_1969,Lemmerz_2025}. As CRs are streaming along the magnetic field down their pressure gradient, they excite Alfv\'en waves, which subsequently undergo collisionless damping and transfer thermal energy to the gas. Here, we emulate CR streaming as a diffusion process (using a diffusion coefficient of $\kappa=0.01$ in code units) and compute the CR-induced Alfv\'en-wave heating via the standard formula (see Eq.~\ref{eq:heat_cr_eqn}). The box is initialized with a uniform gas density of unit magnitude, along with a sinusoidal pressure ($P$) perturbation applied in the $x$-direction:
\begin{align}
P(x,y)   &= P_0 \left(1 + 0.1\sin\left(\frac{2\pi x}{L}\right)\right), 
\end{align}
where $P_0=1$. This setup drives an oscillating sound wave. The CRs are modeled as a separate fluid with their pressure ($P_{\rm cr}$) set to 1\% of the thermal gas, which produces a sinusoidal variation in $P_{\rm cr}$. The velocity is initially zero everywhere, and a uniform magnetic field of unit magnitude is imposed along the $x$-axis, enabling CR transport primarily along the field lines. At the beginning of the simulation, the background plasma starts to oscillate in the longitudinal direction and advects CRs alongside. In addition, the CR pressure gradients drive a flux that diffusively transports CRs relative to the background plasma, thereby softening the CR pressure gradients over time.

In Fig.~\ref{fig:cr_heat_2d_3.5} we compare the time evolution of the total heating rate from the entropy tracer, i.e., $K$ and $s$-based methods, with the estimates obtained by applying the analytical CR heating formula (Eq.~\ref{eq:heat_cr_eqn}) to the simulation output. These curves show excellent agreement at early times. We see quasi-periodic oscillations with a period equal to half of the sound crossing time of the box for the total heating rate, $L/c_\rmn{s} = \sqrt{3/5} = 0.77$. As these sound waves advect CRs along their pressure gradient also changes periodically. Heating occurs in the regions with high CR pressure gradients, causing the CR heating rate to also change periodically. We confirmed that the CR gradients are significantly smoothed after the 1D CR diffusion time $\tau_\rmn{diff}=L_\rmn{cr}^2/(2 \kappa)\sim L^2/(2^5\kappa)\approx3.1$, where we adopted a CR gradient length of $L_\rmn{cr}\sim L/4$, implying weaker Alfv\'en wave heating on this time scale. Indeed, the diffusive relocation of CR energy with time causes the nearly symmetric heating rates at early times to become asymmetric at around $\tau_\rmn{diff}$. Eventually, the heating rate estimated from the entropy-based method starts to deviate from the pure analytical CR heating rate because of additional heating due to numerical viscosity.

\subsection{Dissipative and adiabatic shock heating}
\label{sec:shock_heating_test}
We perform 1D shock-tube tests, using a resolution of 1024 grid points to detect weak and strong shocks and estimate the associated heating rate.

\paragraph{Weak shock:} The simulation domain spans $0.0 \leq x \leq 0.5$, with an initial density of 1 and pressure of 0.1 at $t=0$. The fluid has velocities, $\varv_x = -0.5$ for $x > 0.25$ and $\varv_x = 0.5$ for $x < 0.25$, producing two oppositely propagating shocks with $\mathcal{M}\approx2$.

\paragraph{Strong shock:} The setup is identical except for a lower initial pressure of 0.01 and higher velocities, $\varv_x = -1$ for $x > 0.25$ and $\varv_x = 1$ for $x < 0.25$, resulting in two shocks with $\mathcal{M}\approx 10.4$.

In Fig.~\ref{fig:shock_test}, the left and middle panels show density (top), pressure (middle), and Mach number (bottom) profiles for weak and strong shocks, respectively. The analytical solutions (black dashed lines), derived from the initial gas conditions and shock speed, are shown together with the simulation results (red curves). The agreement in the bottom panels demonstrates that the shock finder accurately recovers the Mach numbers for both shocks. The right panel shows the distribution of dissipative and adiabatic heating within the simulation domain (as described in Sect.~\ref{sec:theory_shock_heating}), evaluated at different times. The results agree well with the theoretical expectations, which are indicated by the dashed vertical lines.

\subsection{Adiabatic heating}
\label{sec:adiabatic_heating_test}

We verify the adiabatic generation of thermal energy using a simple 1D standing sound wave setup, with a resolution of 1024 grid points. 
The initial conditions for the density ($\rho$), pressure ($P$) and velocity ($\bm{\varv}_x$), are given by the analytic solutions:
\begin{align}\label{eq:stand_wave}
\rho(x,t) &= \rho_0 \left[ 1 + A \cos(k x) \cos(\omega t) \right], \notag\\
P(x,t) &= P_0 \left[ 1 + \gamma A \cos(k x) \cos(\omega t) \right],\, {\rm and} \\ \notag
\bm{\varv}_x(x,t) &=  A \, c_s \, \sin(k x) \sin(\omega t) \hat{\bm x},
\end{align}
where $A = 10^{-4}$ is the amplitude of perturbation, $k = 2 \pi n / L$ is the wave number,  $\omega = c_s k$ is the angular frequency, with $c_s$ being the sound speed, and $n$ is the mode number. Since the perturbation amplitude is small ($A \ll 1$), the evolution remains in the linear regime and is dominated by the fundamental mode ($n=1$). We verify that the time evolution of the physical wave variables follows the expected solution presented in Eq.~\eqref{eq:stand_wave}.  

\begin{figure}
\centering
\includegraphics[width=\linewidth, keepaspectratio]{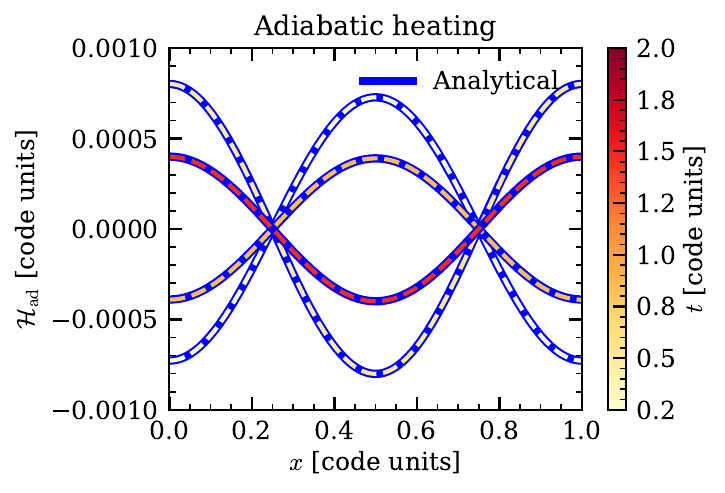}
\caption{Spatial distribution of the adiabatic heating term for a standing wave setup at different times. The simulation output at different times is overplotted with dashed curves over the analytic solution (blue curves).}
\label{fig:ad_heat_test}
\end{figure}

A standing wave produces spatially stationary regions of compression and rarefaction, where the density and pressure oscillate about their equilibrium values ($\rho_0, P_0$). This configuration provides a clean test of the adiabatic heating term from the simulation, since the compressional work, $-P \bnabla \bcdot \bm{\varv}$ 
can be computed analytically from Eq.~\eqref{eq:stand_wave}. In Fig.~\ref{fig:ad_heat_test}, we compare the simulation output of the adiabatic heating with the analytical prediction, which can be approximated as
\be 
  -P \bnabla \bcdot \bm{\varv} \approx -P_0 A \omega \cos(kx) \sin(\omega t),
\en
where we only consider the fundamental mode. The excellent agreement with the entropy-based heating estimate demonstrates that the adiabatic heating term is accurately captured by the \textsc{Arepo} computation of the local velocity divergence.

\section{Heating the ICM by an AGN jet outburst}
\label{sec:jet_test}

Building on the tests presented in the previous section, we apply the heating rate estimator and the shock finder to quantify the heating produced by a single AGN jet episode in galaxy clusters. Here, we consider only the heating associated with shocks, which is determined using the shock-finder algorithm (Sect.~\ref{sec:theory_shock_heating}), and use the entropy-based tracer (Sect.~\ref{sec:theory_equations}) to estimate the heating by turbulent dissipation, mixing, and compressible modes. Additional processes, including CRs (Sect.~\ref{sec:theory_cr_heating}) and Braginskii viscosity (Sect.~\ref{sec:theory_turb_heating}), are reserved for future work in a self-regulated jet model. The present study is intentionally simplified and is primarily designed to validate our heating estimation method for complex scenarios like AGN jets and examine the distribution of the various heating channels within the jets.

In our analysis, the entropy-tracer-based method estimates the grid-scale dissipation of kinetic energy, which we have named \textit{turbulently induced heating}. We acknowledge that this numerical heating channel includes dissipation of vortices emerging from a turbulent cascade, turbulently mixing gas of different temperatures, as well as weakly compressible dissipation of sound waves and very weak shocks. Therefore, our definition of turbulently induced dissipation is not restricted to energy cascades originating at large scales, as is conventionally understood in the literature \citep{troccoli_2026}, but also encompasses several processes as explained. We further characterize the solenoidal and compressible components of turbulent heating using the Helmholtz decomposition. As sound waves and shocks correspond to compressible (i.e., curl-free: $\bnabla \bm{\times} \bm{\varv}_{\rm comp} = \textbf{0}$) motions, while shear-driven turbulence is associated with solenoidal (i.e., divergence-free: $\bnabla \bcdot \bm{\varv}_{\rm sol} = 0$) motions, the Helmholtz decomposition provides a natural separation of these two velocity components. The method is described in detail in Appendix~\ref{sec:helm_decompose}, where we perform the decomposition on $\bm{\varw} = \sqrt{\rho} \bm{\varv}$. The local turbulent heating rate density ($\mathcal{H}_{\rm turb}$) is then partitioned into compressible and solenoidal contributions according to their fractional kinetic energies, i.e.
\begin{align}
 \mathcal{H}_i = \frac{\bm{\varw}_i^2}{\bm{\varw}_{\rm sol}^2 + \bm{\varw}_{\rm comp}^2} \mathcal{H}_{\rm turb},
 \quad\mbox{where}\quad
 i\in\{{\rm sol}, {\rm comp}\}.
\end{align}
In the analysis, we name the corresponding heating rates as `Solenoidal' and `Compressible', respectively.

Several studies have used spectral methods to account for heating through turbulent dissipation \citep{martizzi_2019, li_2025}; however, such methods rely on a number of strong assumptions such as steady state, fluctuations on a homogeneous background, and isotropic turbulence. Those assumptions are not generally satisfied, which limits their suitability as a general-purpose estimator of heating rates. We show in Appendix~\ref{app:heat_spectral_entro} that such spectral methods are in agreement with the employed entropy-tracer-based method under idealized conditions of a turbulent box-setup, when averaged over a few eddy turnover times (thus approaching a steady state). However, in a complex system of an AGN jet in a cluster, they can significantly deviate depending on the dynamical state of the flow. In our future work, we plan to explore filtering-based approaches, as employed in \citet{perrone_2026}, to characterize the turbulent dissipation in our simulations. 

We perform three-dimensional MHD simulations of AGN feedback in an idealized Perseus-like cluster. The initial conditions are similar to  \citet{ehlert_2023}, and we refer the reader to this paper for a detailed description and provide a brief overview below.

\subsection{Description of the setup}

\subsubsection{Cluster initial conditions}
The simulations are performed in a cubic domain of size 5~Mpc. We adopt the radial electron density profile ($n_\rmn{e}$) of the Perseus cluster from \citet{churazov_2003}, rescaled to $h=0.67$, which is given as
\begin{align}
n_\rmn{e} =\; & 46 \times 10^{-3}
\left[ 1 + \left(\frac{r}{60\,{\rm kpc}}\right)^2 \right]^{-1.8} \cc \notag \\
&+ 4.7 \times 10^{-3} 
\left[ 1 + \left(\frac{r}{210\,{\rm kpc}}\right)^2 \right]^{-0.87} \cc.
\end{align}

\begin{table}[t!]
\centering
\begin{ThreePartTable}
\caption{Jet model parameters.}
\begin{tabularx}{0.9\columnwidth}{lXXX}
\hline
Label & $\rho_{\rm jet}$ (g\,cm$^{-3}$) & $\eta $ $  (\rho_{\rm jet}/\rho_{\rm amb,0})$ & Jet-On Phase \\
\hline
very-light & $10^{-29}$ & $10^{-4}$ & 50 Myr \\
light      & $10^{-28}$ & $10^{-3}$ & 50 Myr \\
intermediate      & $10^{-27}$ & $10^{-2}$ & 20 Myr\\
\hline
\end{tabularx}
\begin{tablenotes}
\item[] In each simulation, we fix the injected jet power at $3\times10^{45}$~erg~s$^{-1}$, adopt a target mass $m_{\rm target,gas}=1.5\times10^{6}\,\rmn{M}_\odot$, a jet target volume $V_{\rm target,jet}^{1/3}=0.3$~kpc, a radius of the accretion region $r_{\rm acc}=5$~kpc, a jet radius at the injection $r_{\rm jet} = 1.65$~kpc, a magnetic-to-thermal pressure ratio $X_{B,{\rm ICM}}=0.0125$ and $X_{B,{\rm jet}}=0.1$.
\end{tablenotes}
\label{tab:sim_params}
\end{ThreePartTable}
\end{table}

We assume hydrostatic equilibrium of the ICM and adopt a gravitational potential of the cluster that results from an NFW \citep{Navarro_1996} dark matter density profile ($R_{200,{\rm NFW}}=2$ Mpc, $M_{200,{\rm NFW}}=8\times10^{14} \,\rmn{M}_\odot$, $c=5$) and an isothermal mass distribution of the central galaxy ($R_{200, {\rm ISO}}=15$ kpc, $M_{200, {\rm ISO}}=2.4\times10^{11} \,\rmn{M}_\odot$; \citealt{mathews_2006}). The ICM magnetic field is initialized as a divergence-free Gaussian random field with a Kolmogorov spectrum on scales larger than the injection scale, characterized by a wave number $k_{\rm inj}=37.5^{-1}$ kpc$^{-1}$ and a magnetic-to-thermal pressure ratio in the ICM of $X_{B,{\rm ICM}}=P_{B,{\rm ICM}}/P_{\rm th}=0.0125$ \citep[see Appendix A of][]{ehlert_2018}. We use an HLLD Riemann solver with the Powell 8-wave scheme for magnetic divergence control \citep{pakmor_2013}. Temperature and velocity fluctuations are seeded in the ambient medium as Gaussian random fields with Kolmogorov power spectra. The hydrostatic temperature profile is perturbed multiplicatively using a Gaussian random field in $\delta T/T$, characterized by a dispersion $\sigma_{T}=2$ and mean $\mu_T=1$. The velocity field is initialized with Gaussian random fluctuations having a one-dimensional dispersion $\sigma_\varv=70$ km s$^{-1}$ per component, to mimic substructure formation- and virialization-induced ICM turbulence. Cooling is followed down to $10^4$~K, including primordial and metal-line contributions. We assume a constant metallicity of 0.3~$\rmn{Z}_\odot$, consistent with the observed uniform ICM iron abundance ($Z_{\rm Fe} \approx 0.3~\rmn{Z}_\odot$) within $r\lesssim 1.5$~Mpc \citep{werner_2013}.

\subsubsection{Modeling the AGN jet}
For the jet injection, we adopt the approach described in \citet{weinberger_2023}, and refer the reader to their work for further implementation details. We explain the most important modeling concepts below and summarize the parameters of all simulations in Table~\ref{tab:sim_params}. The jet is launched from within a spherical region of radius $r_{\rm jet} = r_{\rm acc}/3$ located at the cluster center (accretion radius: $r_{\rm acc}=5~\kpc$), where the gas density is set to $\rho_{\rm jet}$. The mass that is removed from (or added to) this jet region is not redistributed to the surroundings, but added to the gravitating mass of the black hole, ensuring total mass conservation. Within this region, the thermal energy is adjusted to bring the cells at least into pressure equilibrium with their surroundings. The remaining available jet energy, $L_{\rm jet}\,\Delta t$ (where $\Delta t$ is the simulation time step), is then injected as kinetic energy in two opposite directions. A passive jet scalar (a proxy for jet mass fraction), $X_{\rm jet}=1$, is initialized in the jet region to trace its subsequent evolution.

In this study, the jets are injected bi-directionally with a $10^\circ$ half-opening angle\footnote{This differs from \citet{weinberger_2023}, where the jets are launched without any opening angle; we find that this difference does not have a significant impact on the heating analysis done in this study.} from the central region. A helical magnetic field is initialized in the jet fluid with a magnetic-to-thermal pressure ratio of $X_{B,{\rm jet}} = 0.1$. In this study, we perform simulations of jets with an equal power of $3\times10^{45}~\ergs$, but varying their density ($\rho_{\rm jet}$) or contrast ($\eta = \rho_{\rm jet}/\rho_{\rm amb,0}$) with respect to the ambient medium at the injection zone. Here $\rho_{\rm amb,0}$ is the density in the cluster's center at $t=0$. We label the runs as very-light, light, and intermediate based on their density contrast.\footnote{We adopt these names because, in a forthcoming study on self-regulated jets, we intend to explore a wider range of jet densities. These labels will thus provide a consistent and intuitive naming scheme spanning from very light to very dense jets.}

The target cell mass in the simulations increases with radius as $m_{\rm target} = m_{\rm target,0}\exp(r/150~{\rm kpc})$. To focus the computational resources on the object of interest, we refine cells with $X_{\rm jet} > 10^{-3}$ down to a target volume $V_{\rm target,jet}$, enforcing a maximum volume ratio of 4 between neighboring cells \citep[see][]{weinberger_2017}. We also apply an additional shock-refinement step, in which cells flagged as being in shock zones are refined to achieve a resolution ten times higher than the target cell mass. 

\begin{figure*}
\centering
\includegraphics[width=0.9\linewidth, keepaspectratio]{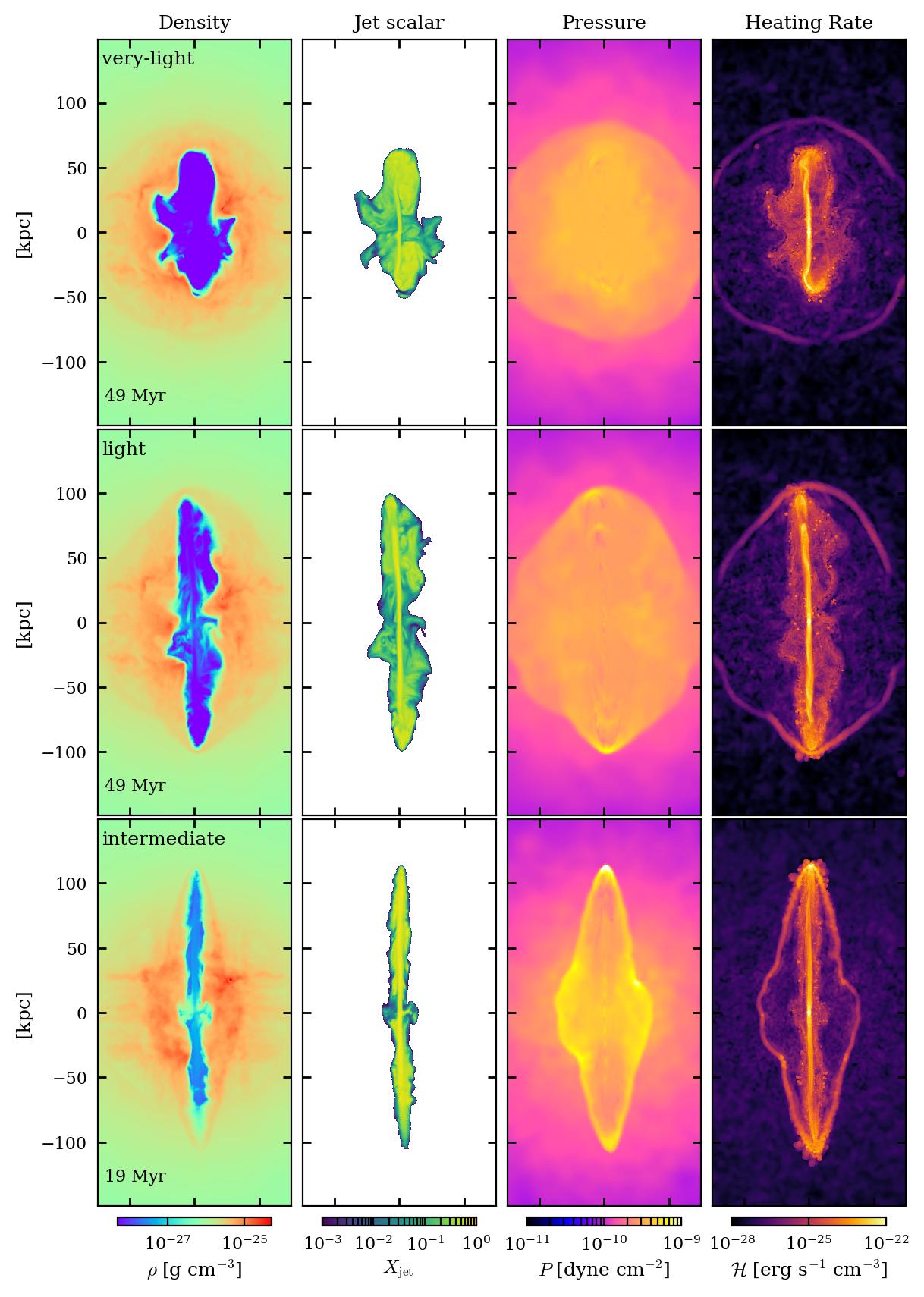}
\caption{Volume-weighted projected density, jet scalar (or jet mass fraction), pressure, and heating rate (5 kpc deep) for the different simulations. Here, the jets are still active and will be turned off after being powered for one more Myr. In the second column, regions with $X_{\rm jet} < 10^{-3}$ are excluded, thereby isolating the jet lobe regions.}
\label{fig:heat_collage_1}
\end{figure*}

\begin{figure*}
\centering
\includegraphics[width=0.9\linewidth, keepaspectratio]{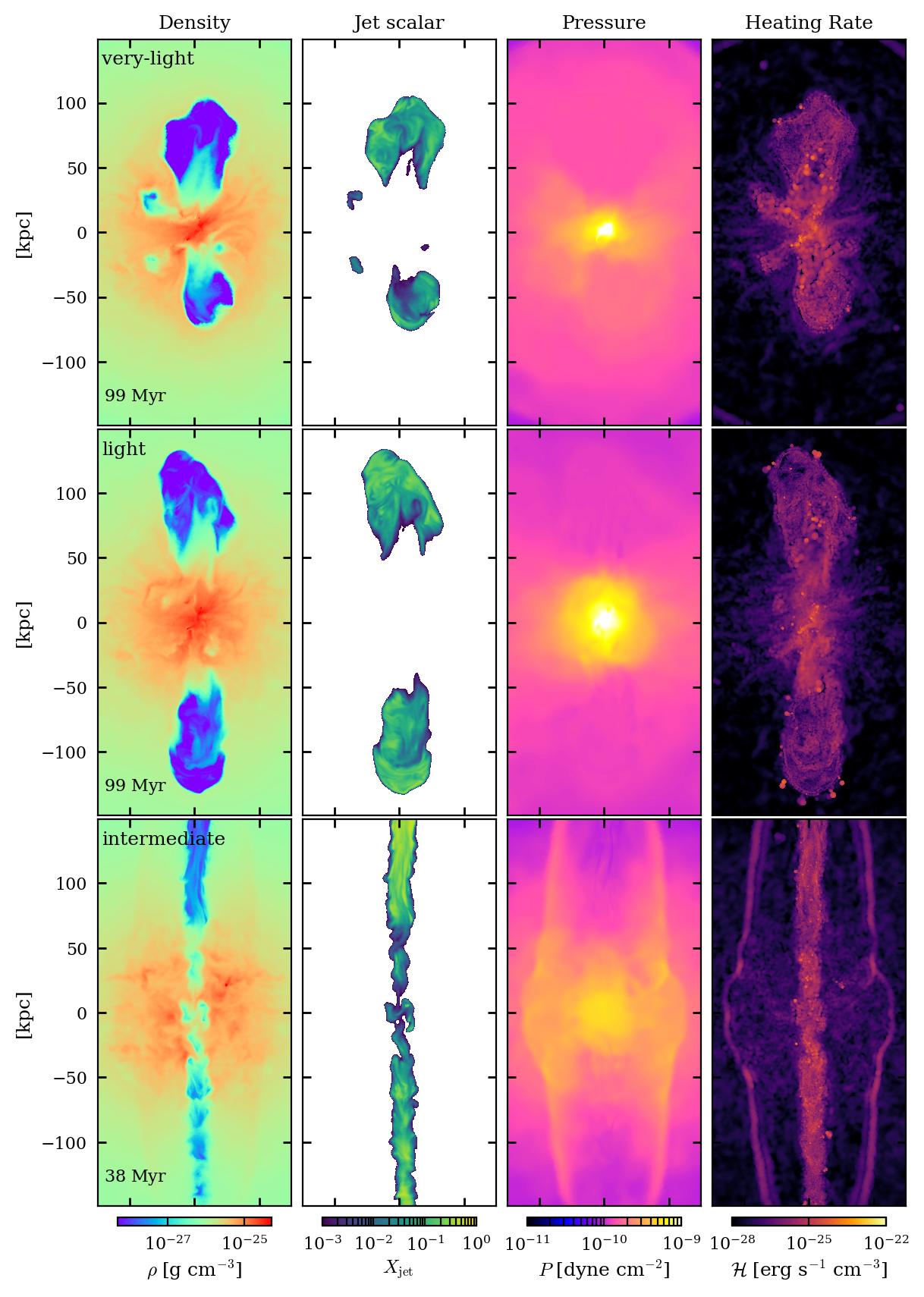}
\caption{Same as Fig.~\ref{fig:heat_collage_1} but shown at later stages when all jets have been turned off. The elapsed time since jet shut-off is 49~Myr for the first two cases and 18~Myr for the bottom row.  In the second column, regions with $X_{\rm jet} < 10^{-3}$ are excluded, thereby isolating the residual jet lobe regions.}
\label{fig:heat_collage_2}
\end{figure*}

\begin{figure*}
\sidecaption
\includegraphics[width=12cm, keepaspectratio]{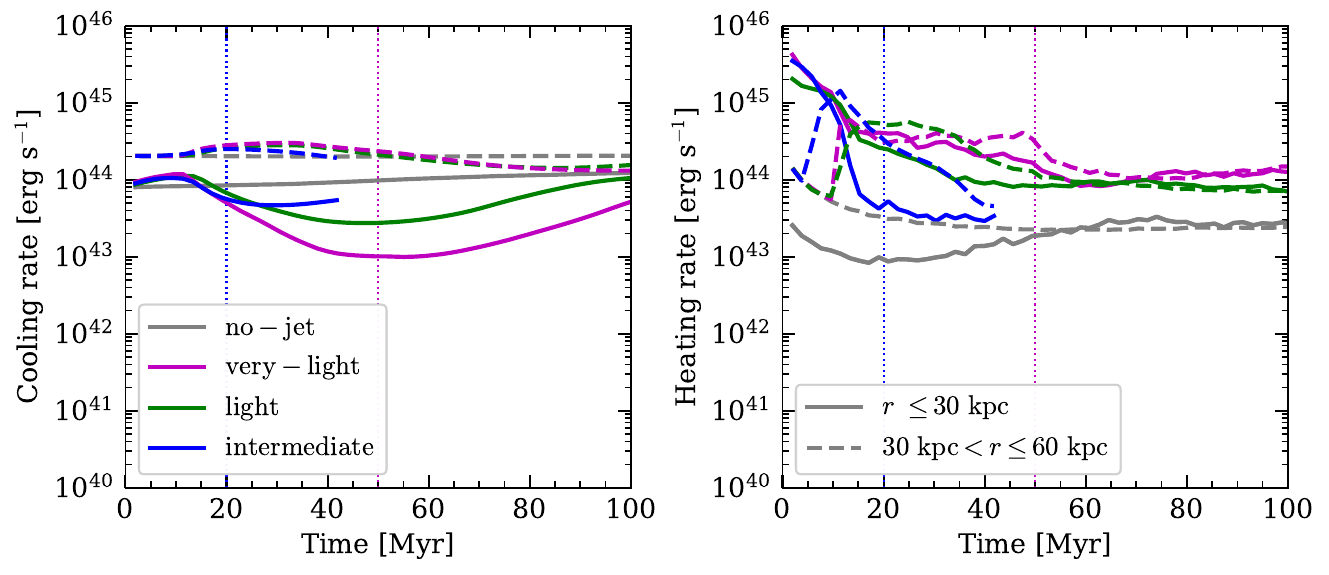} 
\caption{Time evolution of the cooling and heating rates of gas within 30 kpc (solid) and 30--60 kpc (dashed) for different jet cases, including the no-jet case. The cooling rate is estimated for gas with temperatures above 0.1 keV ($T\gtrsim 1.2\times 10^{6}~$K). The vertical dotted lines indicate the jet turn-off times.}
\label{fig:cool_rates}
\end{figure*}

\subsection{Comparative evolution of jets with different densities}
In Figs.~\ref{fig:heat_collage_1} and~\ref{fig:heat_collage_2}, we show projections with a depth of 5~kpc depicting gas density, jet scalar ($X_{\rm jet}$), pressure, and heating rate. From top to bottom, the rows correspond to very-light to intermediate-density jets. In Fig.~\ref{fig:heat_collage_1}, the jets are still active, whereas in Fig.~\ref{fig:heat_collage_2}, they have been turned off for some time. The very-light and light jets are switched off at 50~Myr, by which time the former has reached a height of ${\sim}50~$kpc and the latter ${\sim}100~$kpc. The intermediate-density jet is turned off at 20~Myr, having almost reached a height of $100~$kpc. 

As can be seen from Fig.~\ref{fig:heat_collage_1}, the light jets form wide, quasi-spherical bubbles, enabling them to displace a larger volume of the ICM in the cluster core. In contrast, an intermediate-density jet, carrying higher momentum, propagates more rapidly through the ICM and generates narrower cocoons. Consequently, at comparable spatial scales, denser jets sustain higher pressures within their cocoons and at the bow shock than their lighter counterparts. The heating maps reveal that, in all cases, the jets induce strong heating along them, with additional heating in the surrounding cocoon. The forward shocks of the intermediate-density jet are noticeably stronger than those of light jets (as seen in the pressure maps), leading to more intense heating at the shock surface.

After the jets are switched off (Fig.~\ref{fig:heat_collage_2}), the jet-inflated cavities begin to rise buoyantly. During this phase, mixing between the jet plasma and the surrounding ICM becomes increasingly prominent, as evident from the jet-scalar maps. Heating within the cocoons weakens; nevertheless, moderate heating persists as the residual kinetic energy stored in the cavities gradually dissipates. Additional localized heating arises from mixing between the jet plasma and the ambient ICM. Meanwhile, the forward shocks continue to propagate for some time, producing weak heating in the outer regions. These different heating processes by the jets are expected to influence the thermal evolution of the ICM in the cluster core by affecting its radiative cooling rate, which we will examine next.

\subsubsection{Cooling and heating in the cluster's central region}
In Fig.~\ref{fig:cool_rates}, we show the time evolution of cooling and heating rate of the gas in the central regions of clusters, i.e., for cluster-centric radii $r\leq30$~kpc and $30<r\leq60$~kpc. For clarity, we refer to the central  $r\leq30~$kpc region as the inner core, and the $30-60~$kpc region as the outer core. The gray curve represents a run without a jet, while the colored curves correspond to the different jet cases. In the cooling-rate estimates, we include only gas with temperatures $\geq$~0.1~keV to focus on the gas cooling from the `hot' to the `cool' phase. The total cooling rates include both metal-line and free-free cooling computed in situ, where the latter starts to dominate at high temperatures. For the heating-rate estimates, we limit the analysis to regions outside the jet-dominated plasma ($X_{\rm jet}\leq 0.9$). The total heating rate is estimated by summing the heating estimation from the entropy tracer for non-shocked regions and shock-dissipative heating. We consider only the irreversible component of the heating, as this is the part that generates entropy, and represents the genuine heat added to the system. Adiabatic processes (and corresponding heating) modify the thermal energy and may locally increase or decrease it; however, they are thermodynamically reversible and thus do not contribute to the net heat addition. The vertical dotted lines mark the jet turn-off times, using colors consistent with the respective cooling and heating curves. 

The left panel in Fig.~\ref{fig:cool_rates} shows that the light jets, which inflate much wider cavities than the intermediate-density jet (see Fig.~\ref{fig:heat_collage_1}), are the most effective at suppressing cooling in the inner core when compared to the no-jet run. In contrast, the intermediate-density jet remains largely collimated and narrow, and therefore is unable to offset the cooling efficiently. At larger radii, these jets have only a limited impact, and beyond the core, the cooling rates closely follow those of the no-jet case. Although compression by the jet's forward shock briefly enhances the cooling rate due to the higher-density shocked gas, and the subsequent buoyant rise of the jet-inflated cavities leads to a modest reduction in cooling, these effects remain minor overall.

The heating-rate evolution (right panel in Fig.~\ref{fig:cool_rates}) shows that, in the absence of a jet, the dissipation of pre-existing turbulence is insufficient to offset the cooling losses. While active, the jets in our study deposit heat at levels of ${\sim} 10^{45}\,\ergs$ (comparable to their power) in the central regions. As the jets propagate to large distances, the heating within the inner core declines, as an increasing fraction of the jet energy is deposited farther out. This behavior is most clearly visible for the light (green) and intermediate-density (blue) jets. Once the jets are switched off, the heating estimates drop sharply by roughly an order of magnitude. Nevertheless, these post-outburst heating rates remain above those of the no-jet case, indicating that residual jet-driven heating can persist in the center for several tens of Myr after the jet is switched off, as can also be seen in Fig.~\ref{fig:heat_collage_2}.

\subsection{Characterizing jet-induced heating mechanisms}

\begin{figure*}
\centering
\includegraphics[width=\linewidth, keepaspectratio]{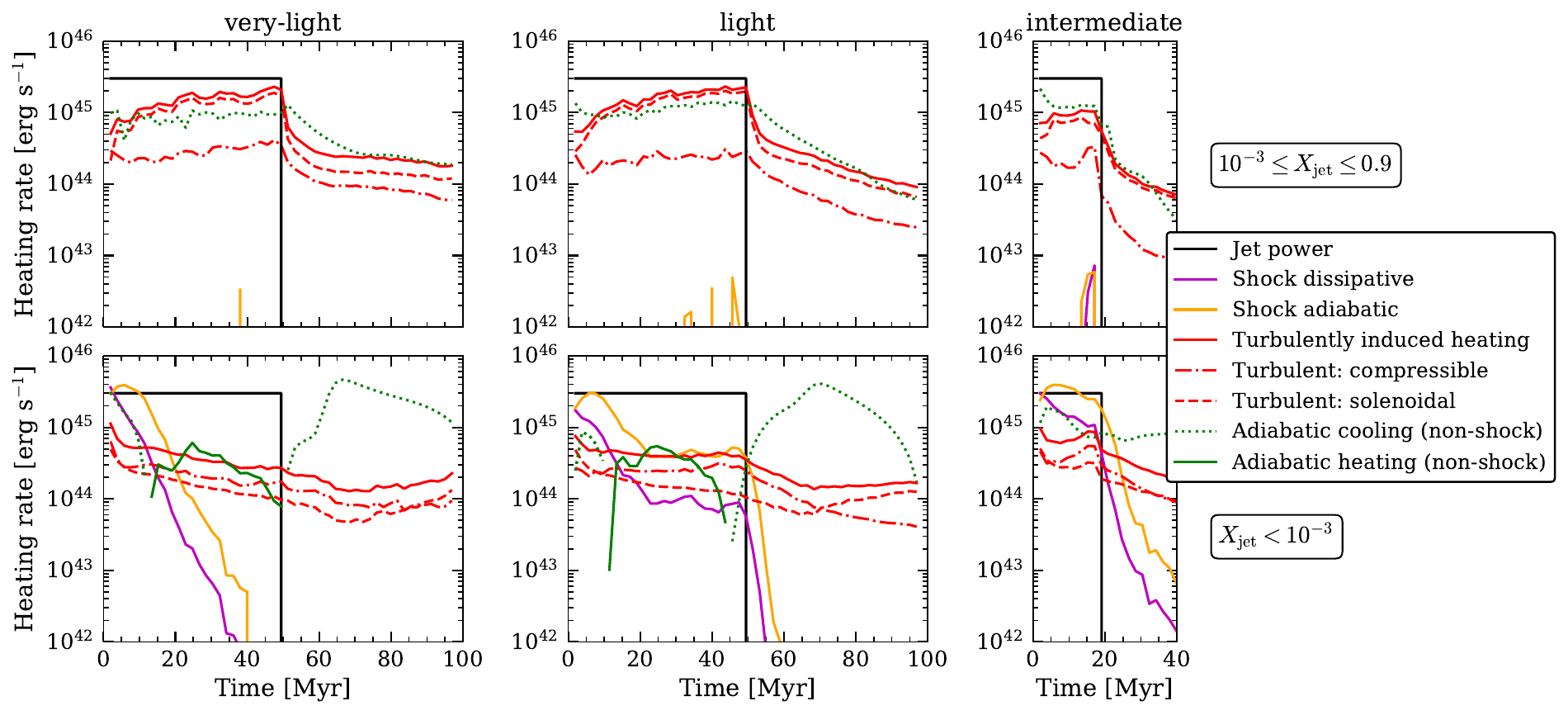}
\caption{Heating rates as a function of time for different jet simulations, within radii $r<100$~kpc. The top row shows the total heating from regions inside the jet cocoon, excluding the jet plasma-dominated region ($X_{\rm jet}\leq 0.9$), while the bottom row corresponds to the ICM outside the cocoon. We refer the reader to Sect.~\ref{sec:jet_test} for an explanation of the processes included in turbulently induced heating.}
\label{fig:jet_inst_heat}
\end{figure*}

In this section, we examine how various physical mechanisms contribute to the total heat deposited in the surroundings by the jets. We present the cumulative heating and instantaneous heating rates in Figs.~\ref{fig:jet_cum_heat} and~\ref{fig:jet_inst_heat}, respectively. These estimates are done in regions with a radius less than 100~kpc, corresponding to the maximum height reached while the jets remain active. Additionally, we only consider cells with jet scalar values $X_{\rm jet}\leq0.9$, thereby excluding jet plasma-dominated regions. This ensures that we measure heating associated with the jet’s interaction with the ambient ICM, rather than the thermalisation processes within the jet spine.

In Fig.~\ref{fig:jet_cum_heat}, the top row displays the cumulative energy injection ratios of kinetic ($E_{\rm kin}$), thermal ($E_{\rm th}$), and magnetic ($E_{\rm mag}$) components as a function of time for different jet models. These curves illustrate how the injected jet energy is partitioned among different forms. The vertical black dashed line indicates the jet turn-off time in each case. For the very-light jet, the thermal energy initially dominates ($t \lesssim 20$~Myr) before reaching approximate equipartition with kinetic energy at later stages. In the light-jet case, kinetic energy becomes dominant after ${\approx}20$~Myr, while for the intermediate-density jet, the kinetic part dominates from the beginning. Overall, with increasing jet density, the energy partition transitions from thermal-dominated to kinetic-dominated injection.  Below, we discuss our inferences on the heating by jets:

\subsubsection{Cumulative heating from different processes}
\begin{figure*}
\centering
\includegraphics[width=\linewidth, keepaspectratio]{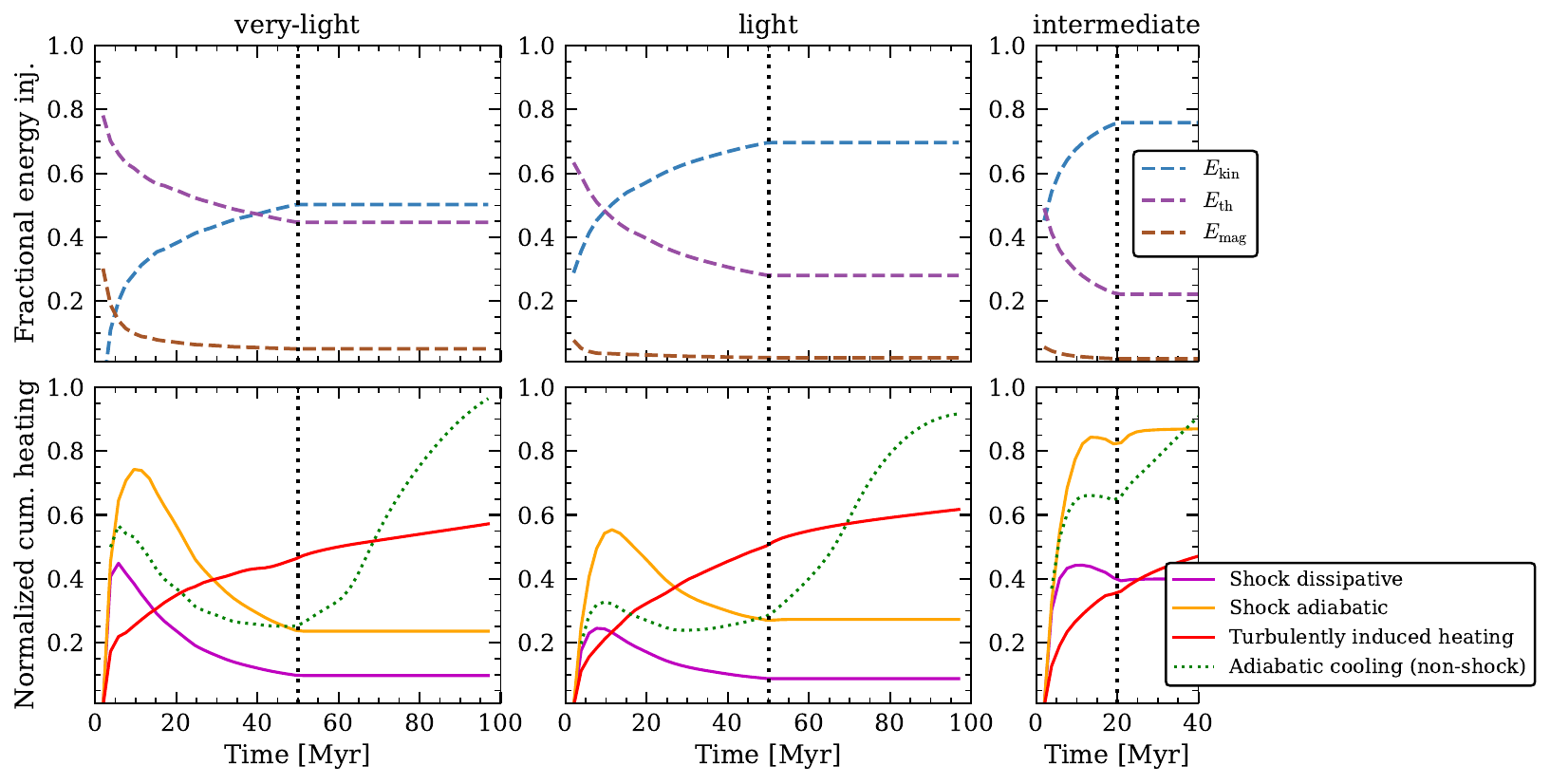}
\caption{Top row: time evolution of the fractional energy injection into kinetic, thermal, and magnetic components in the jet injection region. Bottom row: time-integrated cumulative heating as a function of time, normalized by the total injected energy at each corresponding time, for the different jet simulations. The heating is computed within 100 kpc from the center and for cells with jet scalar $X_{\rm jet}\leq 0.9$. Refer to Sect.~\ref{sec:jet_test} for an explanation of the processes included in turbulently induced heating.}
\label{fig:jet_cum_heat}
\end{figure*}

The bottom row in Fig.~\ref{fig:jet_cum_heat} presents the corresponding time-integrated cumulative heating contributions from shock-dissipative, shock-adiabatic, and turbulent processes, normalized by the cumulative energy injected in the domain at the same time. The adiabatic heating and cooling (for non-shocked regions), representing heating associated with regions undergoing compression, and expansion, respectively, are also shown. While expanding regions correspond to local divergence (lowering of local thermal energy), they can still have positive heating due to viscous and shear dissipation within the turbulent flow, and also possible mixing-induced heating.

\begin{figure*}
\centering
\includegraphics[width=\linewidth, keepaspectratio]{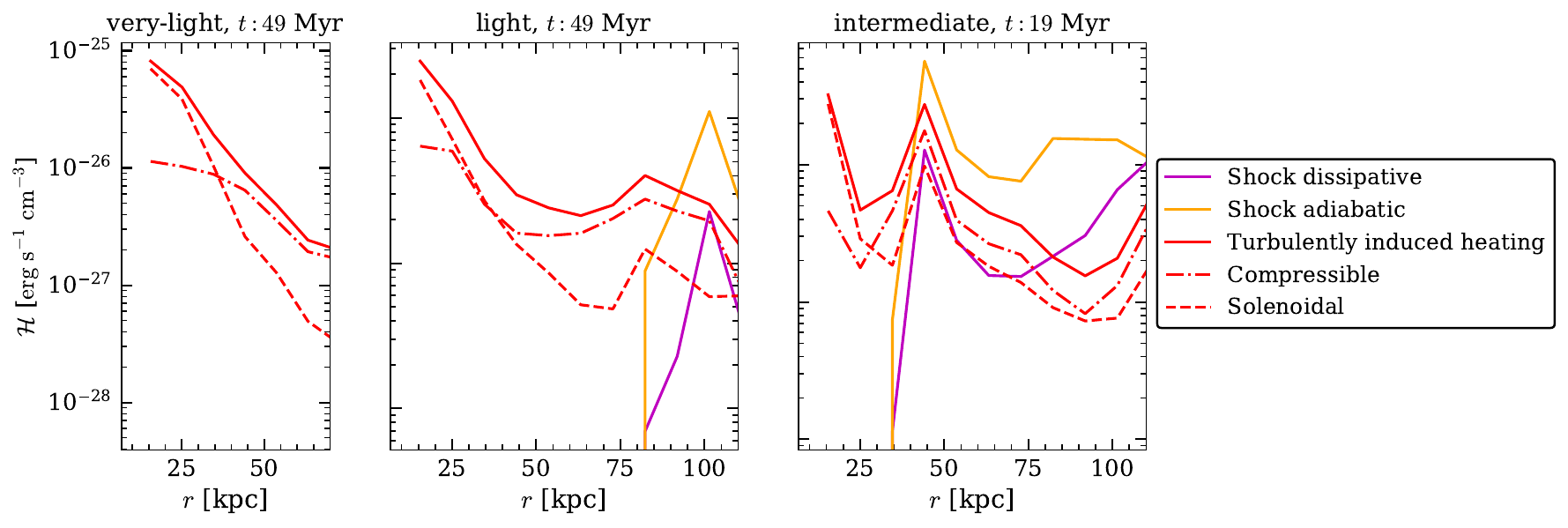}
\caption{Radial profiles showing the ICM heating rate density as a function of radius for different jet simulations just before the jets are shut off. We compute the heating rate density of the ICM (for $X_{\rm jet}\leq 10^{-3}$), i.e., we exclude jet-lobe regions. The radial extent in each case is limited to the maximum height attained by the jet at the corresponding time.}
\label{fig:rad_jet}
\end{figure*}

We find that the shock heating dominates at early times across all jet cases. The stronger contribution from the shock-adiabatic term compared to the dissipative term indicates that most of these shocks are weak ($\mathcal{M}\lesssim 2.3$; as inferred from Fig.~\ref{fig:diss_ad_plot}). For the very-light and light jets, turbulently induced heating and weakly compressible dissipation (labeled with the overarching term \textit{turbulently induced heating}) becomes increasingly important at later times ($t > 20$~Myr), consistent with the development of wide, buoyant bubbles that promote large-scale mixing and thermalization. The energy in shocks decreases continuously, implying that shocks contribute weakly to the heat added by the jets in the surroundings. In contrast, the intermediate-density jet deposits most of its energy through shocks, with relatively minor turbulent dissipation, signifying more localized energy transfer. There is consistently net adiabatic cooling because the jets induce an overall expanding motion.

Interestingly, after the jets are turned off, the adiabatic cooling persists in all the cases, implying a lowering of local thermal energy due to rising and expanding high-pressure cavities. Additionally, the gradual gain in the cumulative turbulently induced heating in all cases indicates thermalization of the residual energy of the jets, either by mixing or turbulent dissipation, as inferred from the right panels in Fig.~\ref{fig:heat_collage_2}.

\subsubsection{Time-resolved heating by jets}
In Fig.~\ref{fig:jet_inst_heat}, we show the time evolution of different heating mechanisms. We have analyzed separately for regions within the jet cocoon ($10^{-3} \leq X_{\rm jet}\leq 0.9$, top row), i.e., excluding the jet plasma-dominated zones, and in the ICM outside of the cocoon ($X_{\rm jet} < 10^{-3}$, bottom row). The jet-scalar maps in Figs.~\ref{fig:heat_collage_1} and~\ref{fig:heat_collage_2} depict the corresponding regions examined here. In our analysis, turbulently induced heating is further decomposed into contributions from compressible and solenoidal motions as discussed at the beginning of Sect.~\ref{sec:jet_test} and detailed in Appendix~\ref{sec:helm_decompose}. The top panel reveals that inside the jet cocoon, turbulently induced heating (solid red curve) dominates across all cases, with the solenoidal motions contributing the most. This suggests that the heating within the cocoon, away from the jet spine ($X_{\rm jet}\leq 0.9$), arises from solenoidal-shear motions and also possible mixing of jet plasma with the ambient gas (as visible from jet-scalar maps in Fig.~\ref{fig:heat_collage_1}).

In contrast, the bottom row shows that outside the cocoon, shock dissipative and shock adiabatic heating (magenta and orange curves) initially dominate when the jets are compact. Over time, the forward shocks of the light jets weaken as their cocoons expand, and shock heating declines sharply. The intermediate-density jet, however, sustains shock-dominated heating throughout the jet’s active phase, implying that its forward shocks remain strong over longer distances. It should be noted that as the forward shocks weaken while propagating to larger distances, the shock finder may fail to detect them due to the minimum Mach number threshold of 1.3 (as discussed in Sect.~\ref{sec:shock_theory}). For large-scale jets, most detected shocks can have very low Mach numbers, often close to unity \citep[see e.g.,][]{li_2017,omoruyi_2025}. Such weak shocks primarily compress the gas with very little dissipative heating, as also evident from Fig.~\ref{fig:diss_ad_plot}. Therefore, in our study, we classify these compressible turbulent regions outside the jet cocoon as acoustic compressions arising from the coherent compressible flow of the forward shock, rather than as bona fide shocks. Notably, the bottom row of Fig.~\ref{fig:jet_inst_heat} shows that, especially during the active phase, the increase in thermal energy outside the jet’s cocoon is dominated by heating from weak shocks and acoustic compressions, with the latter being more prominent for light jets. After the jets are switched off, solenoidal motions also begin to contribute appreciably to the heating in regions of low jet-scalar values.

\subsubsection{Radial distribution of heating mechanisms}
The radial profiles of the ICM heating rate density as a function of radius for different heating mechanisms are shown in Fig.~\ref{fig:rad_jet}. The radial domain is divided into spherical shells of width 10~kpc, and the profiles are shown only up to the maximum height reached by the jet in each snapshot. The heating rates are calculated excluding the jet and lobe regions (i.e.\ for $X_{\rm jet} \leq 10^{-3}$) and at times just before the jet was switched off.

In simulations with increasing jet density, the flow becomes increasingly kinetic-energy dominated rather than thermal-energy dominated (see top row of Fig.~\ref{fig:jet_cum_heat}), and shocks become progressively more important at larger radii. In all of the cases, the compressible motions begin to dominate over the solenoidal motions for distances greater than 25~kpc. In the intermediate-density case, particularly beyond $r \gtrsim 75~\mathrm{kpc}$, i.e., near the jet's head, shock heating becomes dominant over compressible modes. This suggests that most of these shocks have Mach numbers between $1.3-2.3$, as the adiabatic compression contributes more to the thermal energy gain than irreversible shock dissipation (as can be inferred from Fig.~\ref{fig:diss_ad_plot}). These results are consistent with \citet{martizzi_2019}, who showed that in strongly kinetic jets, shocks become increasingly important at large radii, while turbulent dissipation dominates in the central regions. However, their analysis considered full spherical volumes including jet and lobe material. In contrast, in our case the jet-lobe regions are excluded; although, within the central lobes themselves, turbulently induced heating (primarily via solenoidal motions) is expected to be dominating (see top panels in Fig.~\ref{fig:jet_inst_heat}).\\

\subsection{Summary and Discussion}
\label{sec:discussion}
In this work, we investigated the evolution of jets with different densities in a cluster environment and analyzed their associated heating using the numerical methods developed and validated in Sects.~\ref{sec:algorithm} and~\ref{sec:tests}, respectively. Consistent with earlier studies \citep{ehlert_2023, tsai_2025}, we find that low-density jets inflate broader cocoons and evolve more slowly. In our simulations, pre-existing turbulence within the cluster's core alone is unable to compensate for the cooling losses in the cluster core, underscoring the need for additional heating sources such as jet feedback, as also emphasized by previous studies \citep[e.g.][]{li_2025}. For a fixed jet power, light jets more effectively offset the cooling losses in comparison to denser jets. This arises mainly because light jets displace a larger volume of central gas, whereas denser jets channel their impact along a narrow path. Importantly, the heating contributed by the light jets persists for several tens of Myr after the jets are switched off, leaving the core with residual yet non-negligible heating (see Figs.~\ref{fig:heat_collage_2} and~\ref{fig:cool_rates}). 

We find that the distribution of different heating processes also varies with the jet's density. For intermediate-density jets, most of the heating arises from shocks driven into the surroundings, whereas the very light jets primarily heat the gas through turbulence and/or mixing between the jet plasma and ambient gas. This aligns with findings of \citet{perucho_2014}, who found shock heating to be important for high-power jets. These high-power jets in their study (with powers of $10^{45}-10^{46}~\ergs$) are highly relativistic, and in some cases, their densities are enhanced to achieve the target power. Consequently, they carry high momentum, similar to the dense jets in our study, making shock heating important on large scales. The jets in our study also transition from thermal-dominated to kinetic-dominated with increasing jet density. The very light jets in our setup retain a significant thermal component throughout the simulation and exhibit predominantly turbulently induced heating over time, consistent with the conclusions of \citet{martizzi_2019}. This result also suggests that the jet’s internal energy composition can play a key role in determining which heating mechanisms dominate.

Consistent with previous simulations \citep{hillel_2016, yang_2016a}, we also find that cocoon regions containing little jet plasma exhibit substantial heating. This heating is expected to have an important contribution due to mixing between the jet material and the ambient gas. In addition, this contribution can also arise from the turbulent dissipation within the cocoon (see Fig.~\ref{fig:decompose_ldj}), an effect considered subdominant in those earlier works. This is because the tracer-based analyses in their studies showed that gas parcels within the cocoon gain thermal energy without a corresponding increase in kinetic energy. In reality, mixing, turbulence, and backflow-induced compression can have a complex interplay that governs the gas’s thermal evolution within the cocoon. The entropy generation observed in our simulations suggests that turbulent dissipation plays a significant role, although isolating its contribution from mixing remains challenging. 

Outside the cocoon, heating is dominated by weak shocks. As the jet propagates and expands, these shocks progressively weaken, reaching Mach numbers close to unity \citep{li_2017, ehlert_2018, omoruyi_2025}, and behave like acoustic compressions.

\section{Conclusions}
\label{sec:conclusions}
 We studied the energy dissipation of a single AGN jet outburst in a galaxy cluster environment. To this end, we have implemented a heating estimation method in \textsc{Arepo} by tracking entropy tracers, specifically $P\rho^{-\gamma}$ and $\log (P\rho^{-\gamma})$, as passive scalars (see Sect.~\ref{sec:theory}). To check its validity, we have performed tests for different heating mechanisms (Sect .~\ref {sec:tests}), particularly for turbulent dissipation, Braginskii viscosity heating, and CR Alfv\'en wave heating. We have also validated the shock-finder for detecting shocks and estimating associated heat input into the gas. In all of the cases, we have found good agreement with the analytical results. We used the heating estimator and shock finder to quantify the turbulent/mixing heating and shock heating produced by fixed-power jets with varying densities (Sect.~\ref{sec:jet_test}). Furthermore, we used Helmholtz decomposition (Appendix~\ref{sec:helm_decompose}) to separate the turbulently induced heating caused by compressible and solenoidal modes. Below, we list our key findings on the jet heating mechanisms:
\begin{itemize}
    \item Our results indicate that pre-existing turbulence cannot balance the cooling losses in the cluster core, highlighting the crucial role of AGN feedback, particularly jets. We found that, for a fixed high-power outburst, light jets can offset the cooling losses far more efficiently than intermediate-density jets. Furthermore, the residual heating from the jets persists for several tens of Myr, even after the jets are switched off (Fig.~\ref{fig:heat_collage_2}).
    
    \item The light jets are able to produce wider bubbles when compared to the intermediate-density jet (see Fig.~\ref{fig:heat_collage_1}). This indicates that these light jets can displace a large fraction of gas in the central parts of the clusters, whereas intermediate-density jets propagate easily to large distances without affecting a large fraction of the volume of gas in the cluster's core.

    \item In our study, the light jets are initially thermally dominated, and the kinetic components strengthen gradually with time, whereas intermediate-density jets are mostly kinetically dominated from the start. The mode of the cumulative energy transfer to the surrounding medium differs accordingly: denser jets inject most of their energy through shocks, while lighter jets primarily do so through turbulently induced heating (which includes dissipation of vortices emerging from a turbulent cascade and mixing gas of different temperatures) and compressible dissipation of sound waves and very weak shocks. However, at the early stages of jet evolution, shock heating dominates for both light and intermediate-density jets (inferred from Fig.~\ref{fig:jet_cum_heat}).

    \item We find that inside the jet cocoon, heating is dominated by turbulent dissipation of solenoidal-shear modes and also potential mixing in all of the cases. Outside the cocoon, shocks remain strong in the intermediate-density jets but weaken rapidly in the light jets. In the latter, most of the heating arises from compressible modes such as shocks and acoustic compressions (see Fig.~\ref{fig:jet_inst_heat}).
\end{itemize}

In this work, we have restricted our analysis to a single jet-outburst with a fixed power in the galaxy clusters. In realistic cluster environments, however, a cooling flow develops over time toward the core, and the jet power becomes self-regulated through accretion onto the central black hole. This feedback can affect both the jet dynamics and the overall heating efficiency. For example, light jets can struggle to maintain a steady power output and to propagate coherently through the surrounding ICM; they are more easily disrupted, deflected, or decelerated by ambient gas motions. In contrast, jets with high densities can traverse the cluster environment more easily, retaining their momentum over longer distances. We will explore these effects in detail in a forthcoming paper, where we will examine the evolution of self-regulated jets in clusters and the heating mechanisms associated with their long-term feedback cycles.

\begin{acknowledgements}
     We thank the anonymous referee for their valuable comments, which helped in improving the clarity of the paper.
      MM and CP acknowledge support by the European Research Council under ERC-AdG grant PICOGAL-101019746, and support by the DFG Research Unit FOR-5195. RW acknowledges funding of a Leibniz Junior Research Group (project number J131/2022). TB gratefully acknowledges funding from the European Union’s Horizon Europe research and innovation programme under the Marie Skłodowska-Curie grant agreement No 101106080 and financial support by the Carlsberg Foundation via grant CF23-0417. The authors gratefully acknowledge the computing time granted by the Resource Allocation Board and provided on the supercomputer Emmy/Grete at NHR-Nord@Göttingen as part of the NHR infrastructure. The calculations for this research were conducted with computing resources under the project JARDS ID 26548, and using the in-house computing facility `Newton' at AIP.
      The \textsc{Arepo} data were processed using \textsc{Paicos} \citep{Berlok2024}, which relies on \textsc{Astropy} \citep{Astropy2013} for unit handling and incorporates parallelized routines implemented in \textsc{Cython} \citep{Cython2011}. Plots were generated with \textsc{Matplotlib} \citep{Hunter2007}. 
\end{acknowledgements}

\bibliographystyle{aa}
\bibliography{references}

\appendix

\section{Finite-volume discretization issues}
\label{app:discretization_issue}

In this section, we demonstrate how discretization in finite-volume methods can give rise to spurious (or non-physical) heating rates. In particular, we show that the $K$- and $s$-based diagnostics can diverge and even yield non-physical negative values, primarily due to the advection of contact discontinuities with entropy jumps in combination with discretization errors on the mesh (see Sect.~\ref{sec:post_contact}) and due to reconstruction errors (Sect.~\ref{sec:jet_first_order}).

The thermal energy density can be estimated using entropy tracers (defined in Sect.~\ref{sec:numerical_approach}) via
\begin{align}
    \eps = \frac{P}{\gamma-1} = \frac{\rho^\gamma K}{\gamma - 1} = K_0 \frac{\rho^\gamma}{\gamma-1} \exp\left(\frac{s}{c_V}\right)
    \label{eq:eth}
\end{align}
where the different physical variables have been defined in Sect.~\ref{sec:theory_equations}. For the hydrodynamics case, the differential form of the Euler equations in the absence of explicit heating and cooling terms is \citep{Weinberger2020}
\begin{align}
    \frac{\partial \rho}{\partial t} & + \bnabla\bcdot\left( \rho \bm{\varv}\right)  = 0\\
    \frac{\partial \rho \bm{\varv}}{\partial t} & + \bnabla\bcdot\left( \rho \bm{\varv} \bm{\varv} + P \mathcal{I}\right) = 0 \\
    \frac{\partial \mathcal{E}}{\partial t} & + \bnabla \bcdot \left[ \bm{\varv} \left(\mathcal{E} + P\right)\right] = 0
\end{align}
where $\mathcal{E} = \eps + \frac{1}{2} \rho {\varv}^2$ is the total energy density.


In the finite-volume discretization, the conservative equations for mass ($m$), momentum ($\bm{p}$), and energy ($E$) of a computational cell that is bounded by faces are solved by the following equations:
\begin{align}
    \frac{{\partial} m}{{\partial} t} &= -\sum\limits_{i\, \in\, \mathrm{faces} } (\rho \bm{\varv})_i \bcdot \bm{A}_i \\
    \frac{{\partial} \bm{p}}{\partial t} &= -\sum\limits_{i\, \in\, \mathrm{faces} } (\rho \bm{\varv} \bm{\varv} + P\mathcal{
I})_i \bcdot \bm{A}_i \\
    \frac{{\partial} E}{{\partial}t} &= -\sum\limits_{i\, \in\, \mathrm{faces} } \left(\frac{\rho {\varv}^2}{2}  + \eps + P\right)_i \bm{\varv}_i \bcdot \bm{A}_i
\end{align}
where $\bm{A}_i$ denotes the area vector of each face $i$. An equivalent discretization for $K$ or $s$, from the entropy conservation equation (in the absence of heating/cooling, i.e., Eq.~\ref{eq:cont_eq}), is 
\begin{align}
    \frac{{\partial}(mK)}{{\partial}t} &= -\sum\limits_{i\, \in\, \mathrm{faces}} \left( K \rho \bm{\varv} \right)_i \bcdot \bm{A}_i, \\
    \frac{{\partial}(ms)}{{\partial}t} &= -\sum\limits_{i\, \in\, \mathrm{faces}} \left( s \rho \bm{\varv} \right)_i \bcdot \bm{A}_i .
\end{align}
In the section below, we demonstrate how the entropy conservation on a discrete grid can cause discrepancies in the thermal energy estimation at contact discontinuities.

\subsection{Total energy vs.\ entropy equations across a contact discontinuity}
\label{sec:post_contact}
We consider a contact discontinuity propagating from left to right across an interface of area 
$A$ separating two computational cells in one dimension. The states in the cells to the left and right of the interface are denoted by the subscripts $\mathrm{L}$ and $\mathrm{R}$, respectively. We focus on the evolution of the state variables in the cell to the right after a single explicit Euler timestep, which initially has state variables density, velocity, energy, and pressure $\rho_0, {\varv}_0, E_0$, and $P_0$, respectively.

We consider the case of $\varv_\mathrm{L}=\varv_\mathrm{R}=\varv_0>0$, $P_\mathrm{L}=P_\mathrm{R}=P_0$ but $\rho_\mathrm{L} \neq \rho_\mathrm{R}$. The change in conserved variables of the cell on the right-hand side after a single timestep $\Delta t$ is given by
\begin{equation}\label{eq:updates}
\begin{aligned}
    m   &= m_0 + (\rho_\mathrm{L} - \rho_\mathrm{R})\, \Delta V, \\ 
    {p} &= p_0 + (\rho_\mathrm{L} - \rho_\mathrm{R})\, \varv_0\,\Delta V,  \\
    E  &= E_0 + \tfrac{1}{2} (\rho_\mathrm{L} - \rho_\mathrm{R}) \varv_0^2\, \Delta V ,\\
    mK  &= m_0 K_0 + (\rho_\mathrm{L}K_\mathrm{L} - \rho_\mathrm{R} K_\mathrm{R})\, \Delta V,\\
    ms  &= m_0 s_0 + (\rho_\mathrm{L} s_\mathrm{L} - \rho_\mathrm{R} s_\mathrm{R})\, \Delta V,
\end{aligned}
\end{equation}
where we defined the volume element that is advected across the interface in one timestep as $\Delta V=\varv_0 A\,\Delta t$. Note that the choice $P_\mathrm{L}=P_\mathrm{R}=P_0$ implies $\eps_\mathrm{L}=\eps_\mathrm{R}=\eps_0$ and thus a vanishing enthalpy flux.

\begin{itemize}
    \item {Thermal energy estimation via the energy $E$:}

In the finite volume solver, the thermal energy is recovered from the total energy equation after subtracting the kinetic energy, which is recovered from the continuity and momentum equations. Thus,
\begin{align}
\eps V
&= E - \frac{{p}^{\,2}}{2m} \nonumber\\
&= E_0 + \tfrac{1}{2}(\rho_\mathrm{L} - \rho_\mathrm{R})\,\varv_0^2\, \Delta V
      - \dfrac{\left[p_0 + (\rho_\mathrm{L} - \rho_\mathrm{R})\,\varv_0^2\, \Delta V\right]^2}
              {2\left[m_0 + (\rho_\mathrm{L} - \rho_\mathrm{R})\,\Delta V\right]} \nonumber\\[4pt]
&= E_0 + \tfrac{1}{2}(\rho_\mathrm{L} - \rho_\mathrm{R})\,\varv_0^2\, \Delta V
      - \dfrac{\varv_0^2\left[m_0 + (\rho_\mathrm{L} - \rho_\mathrm{R})\, \Delta V\right]^2}
              {2\left[m_0 + (\rho_\mathrm{L} - \rho_\mathrm{R})\, \Delta V\right]} \nonumber\\
&= E_0 - \tfrac{1}{2}m_0 \varv_0^2
   = \varepsilon_{0} V.
\label{eq:eth2}
\end{align}
From the energy equation, it follows that $\eps = \varepsilon_{0}$ since only the kinetic energy changes within the cell. This is a direct consequence of the constant pressure across a contact discontinuity.\\

\item {Thermal energy estimation via the entropic function $K$:}

Combining Eqs.~(\ref{eq:eth}) and~(\ref{eq:updates}), we can calculate the thermal energy after a single timestep from the evolution of $K$,  $\varepsilon_K$.
\begin{align}\label{eq:eth_K}
&\eps_{K} =  \nonumber\\
&\phantom{=~~} \frac{\rho^\gamma K}{\gamma - 1}
= \frac{m^{\gamma-1} m K}{V^{\gamma} (\gamma - 1)} \\
&= \frac{\left[m_0 + (\rho_\mathrm{L}-\rho_\mathrm{R})\,\Delta V \right]^{\gamma-1}
      \left[m_0 K_0 + \left(\rho_\mathrm{L} K_\mathrm{L} - \rho_\mathrm{R} K_\mathrm{R}\right) \, \Delta V \right]}
      {V^{\gamma} (\gamma-1)} \nonumber\\
&= \frac{\left[m_0 + (\rho_\mathrm{L}-\rho_\mathrm{R}) \, \Delta V\right]^{\gamma-1}
      \left[m_0 K_0 + P_0\left(\rho_\mathrm{L}^{1-\gamma} - \rho_\mathrm{R}^{1-\gamma}\right) \, \Delta V \right]}
      {V^{\gamma} (\gamma-1)}  \nonumber \\
&= \frac{\left[\frac{m_0}{\rho_\mathrm{L}} + \left(1-\frac{\rho_\mathrm{R}}{\rho_\mathrm{L}}\right) \Delta V\right]^{\gamma-1}
      \left[m_0 K_0 \rho_\mathrm{L}^{\gamma -1} + P_0\left(1 - \left(\frac{\rho_\mathrm{R}}{\rho_\mathrm{L}}\right)^{1-\gamma}\right) \Delta V \right]}
      {V^{\gamma} (\gamma-1)}  \nonumber \\
&\neq \frac{m_0^{\gamma-1} m_0 K_0}{V^{\gamma} (\gamma - 1)} = \eps_0  \nonumber
\end{align}
The inequality is true for all $\delta\rho \equiv \rho_\mathrm{R} - \rho_\mathrm{L} \neq 0$ (or $\rho_{\rm R}/\rho_{\rm L}\neq 1$), i.e., for actual contact discontinuities. Note that we used the constant pressure to substitute
\begin{align}
    K_\mathrm{L} &= P_0 \rho_\mathrm{L}^{-\gamma}
    \quad\mbox{and}\notag \\
    K_\mathrm{R} &= P_0 \rho_\mathrm{R}^{-\gamma} \notag
\end{align}
for obtaining an equation for $\varepsilon_K$ that depends on $\rho_\mathrm{L}$ and $\rho_\mathrm{R}$.\\

\item {Thermal energy estimation via the entropy $s$:}

Using Eq.~\eqref{eq:eth}, we can also express the thermal energy density as a function of $\rho$ and $s$:
\begin{align}\label{eq:eth_S}
\eps_{s} &= K_0 \frac{\rho^\gamma}{\gamma-1}\,
\exp\!\left(\frac{s}{c_V}\right) \\[4pt]
&= K_0 \frac{m^\gamma}{V^\gamma(\gamma-1)}\,
\exp\!\left(\frac{ms}{mc_V}\right) \nonumber\\
&= K_0 \frac{(m_0 + (\rho_\mathrm{L} - \rho_\mathrm{R}) \, \Delta V)^\gamma}{V^\gamma (\gamma-1)} \nonumber \\
&\quad \times\exp\left( \frac{m_0s_0 + (\rho_\mathrm{L} s_\mathrm{L}-\rho_\mathrm{R} s_\mathrm{R})\, \Delta V}{\left[m_0 + (\rho_\mathrm{L} -\rho_\mathrm{R}) \, \Delta V\right]c_V}\right)  \nonumber \\
&= K_0 \rho_{\rm L}^\gamma\frac{\left[\frac{m_0}{\rho_\mathrm{L}}  + \left(1 - \frac{\rho_\mathrm{R}}{\rho_\mathrm{L}}\right) \, \Delta V\right]^\gamma}{V^\gamma (\gamma-1)} \nonumber \\
&\quad \times\exp\left( \frac{\frac{m_0s_0}{\rho_\mathrm{L}} + s_\mathrm{L}\left(1-\frac{s_\mathrm{R}}{s_\mathrm{L}}\frac{\rho_\mathrm{R}}{\rho_\mathrm{L}}\right)\, \Delta V}{\left[\frac{m_0}{\rho_\mathrm{L}} + \left(1-\frac{\rho_\mathrm{R}}{\rho_\mathrm{L}}\right) \, \Delta V\right]c_V}\right)  \nonumber \\
&\neq K_0 \frac{m_0^\gamma}{V^\gamma(\gamma-1)}\,
\exp\!\left(\frac{m_0 s_0}{m_0 c_V}\right) = \eps_0  \nonumber
\end{align}
where one can again use Eq.~\eqref{eq:updates} for $m$ and $s$ in terms of the L and R states, and employ pressure equilibrium:
\begin{align}
    s_\mathrm{L} &=  c_V \log\left(P_0 \rho_\mathrm{L}^{-\gamma}/K_0\right), \notag \\
    s_\mathrm{R} &= c_V \log\left(P_0 \rho_\mathrm{R}^{-\gamma}/K_0\right). \notag
\end{align}
\end{itemize}
It becomes clear from the above considerations that the three different approaches to recovering thermal energy are not mathematically equivalent. To understand the behavior, we adopt a concrete example (with the variables measured in internal units):
\begin{equation}
\label{eq:params}
\begin{aligned}
    \gamma &= 5/3, \\
    P_0 &= 5/3, \\
    \rho_\mathrm{L} &= 1, \\
    \rho_\mathrm{R} &\in [10^{-2},1], \\
    \Delta V &= 0.01, \, {\rm and }\\
    V &= 1.
\end{aligned}
\end{equation}

\begin{figure}
\centering
\includegraphics[width=0.83\linewidth, keepaspectratio]{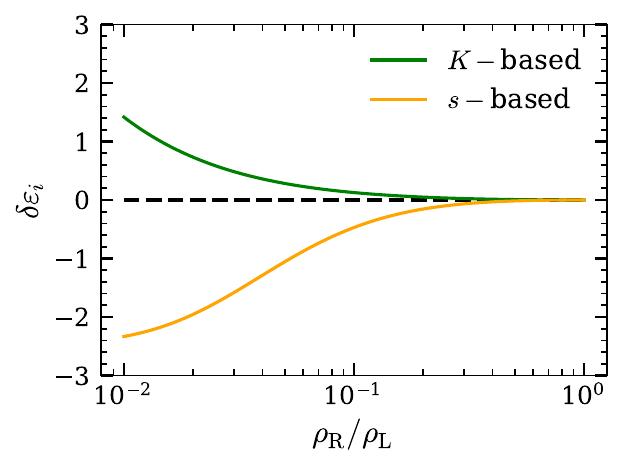}

\caption{Difference in thermal energy estimates ($\delta\varepsilon_i = \varepsilon_i - \varepsilon$, where $i\in\{s,K\}$ and $\eps = 5/2$) using energy (dashed) and entropy (green and orange curves) conservation approach as a function of density jump across the contact discontinuity $(\rho_\mathrm{R}/\rho_\mathrm{L})$. The expressions for $\varepsilon$, $\varepsilon_{K}$ and $\varepsilon_{s}$ can be found in Eqs.~\eqref{eq:eth2}, \eqref{eq:eth_K}, and \eqref{eq:eth_S}, respectively. The two different entropy tracer-based methods can yield deviations near sharp contact discontinuities.}
\label{fig:egy_post_contact}
\end{figure}

We show the difference of the energy density estimated via one of the entropy equations and directly via the energy in Fig.~\ref{fig:egy_post_contact}. All three methods agree at $\rho_\mathrm{R}/\rho_\mathrm{L}=1$, i.e., in the absence of a density jump. However, the $K$-based estimate tends to overestimate the thermal energy, diverging at large density ratios, while the $s$-based estimate tends to underestimate it. Consequently, a heating estimator based on $K$ may indicate spurious negative heating rates near contact discontinuities, whereas an $s$-based estimator would spuriously overpredict the heating rate in these regions. As is evident from the $E$-based reconstruction, such a contact discontinuity should not cause any thermalization. Note that we regularly encounter contact discontinuities with density jumps across four orders of magnitude at the jet-background interface, so that these considerations are critical for accurately capturing the heating rate estimates.

We thus use the diverging heating estimates to identify spurious numerical heating at contact discontinuities and discard the corresponding heating estimates as unphysical. While not shown here explicitly, we would expect similar effects at shocks, where thermalization does occur. In this case, however, we can discard the scalar-based heating estimates as we replace them with the shock finder that calculates the dissipation via shock jump conditions, as discussed in Sect.~\ref{sec:theory_shock_heating}.

\begin{figure*}
\centering
\includegraphics[width=0.93\linewidth, keepaspectratio]{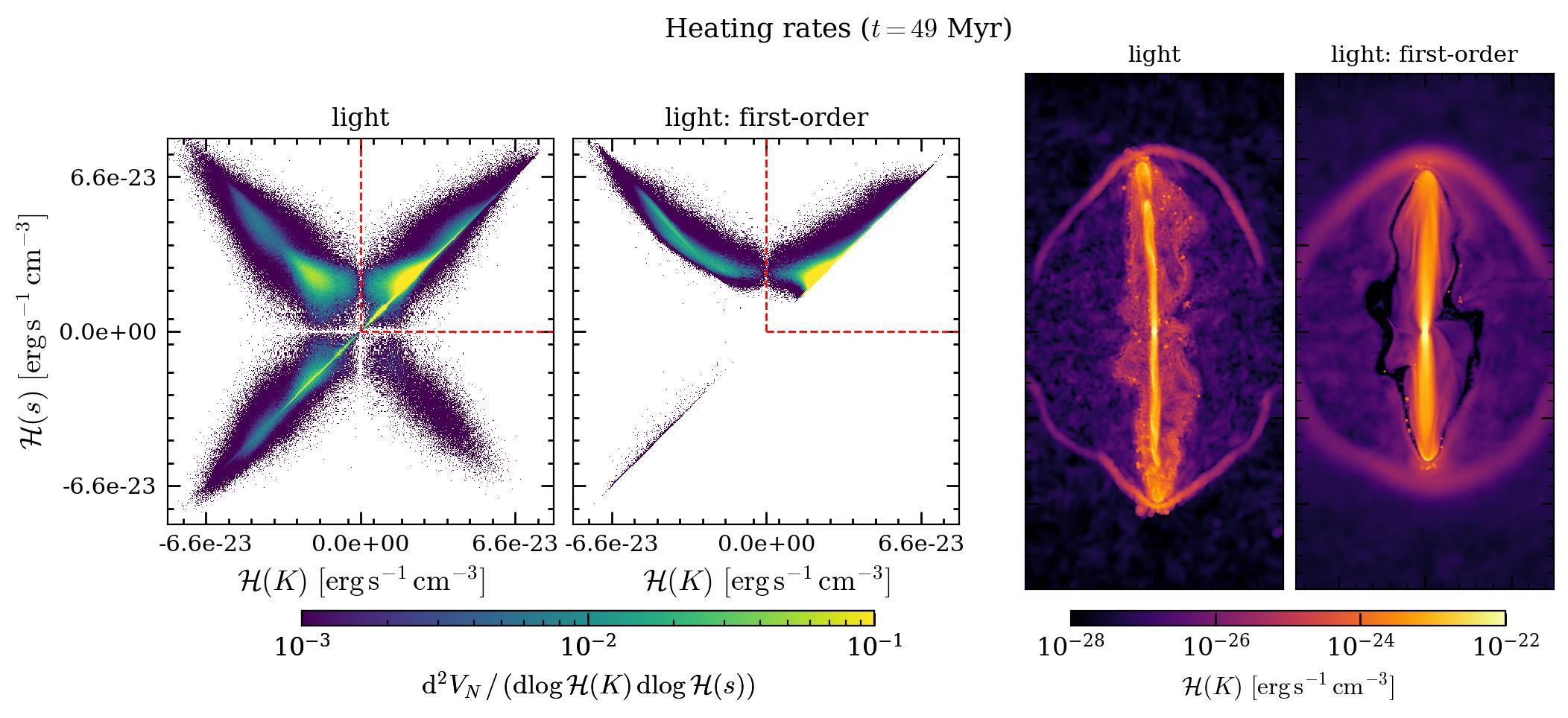}
\caption{Left: Two-dimensional histogram of cell-wise heating rates estimated using the $K$- and $s$-based definitions. Colors denote the volume fraction per logarithmic bin width. The distributions correspond to simulations using the standard approach and ``diffusive configuration'' with only first-order reconstruction of the hydrodynamic quantities for the light jet. The first quadrant, outlined by red dotted lines, marks the region where both methods yield positive heating, indicating genuine heat addition.
Right: Volume-weighted projected heating rates for a 5~kpc-deep projection (see Fig.~\ref{fig:heat_collage_1}), shown for the same two cases as in the left panel.}
\label{fig:heating_K_S}
\end{figure*}

\subsection{Discretization errors in heating estimates in a jet simulation}
\label{sec:jet_first_order}

In this section, we show that reconstruction errors inherent to discretization in finite-volume methods can also introduce non-physical heating in the estimates. In particular, the extrapolated interface entropy obtained from interpolated density and pressure can differ substantially from the interpolated entropy scalar at several locations. Consequently, in a more diffuse scheme, these discrepancies and the resulting non-physical heating are expected to be reduced. We explore these effects in the following analysis.

To assess the impact of discretization on entropy tracer-based heating estimates in the jet simulations, we analyze the distribution of heating rates across different numerical configurations for the light jet case. We compare the estimates from the standard run (light) to the more ``diffuse configuration'' with first-order reconstruction\footnote{In our ``diffuse configuration'', we disabled spatial and temporal extrapolations, turned off scalar gradients, limited velocity and sound-speed slopes during reconstruction, and used the MINMOD slope limiter.}, effectively enforcing a first-order reconstruction scheme (named as light: first-order). The two-dimensional (2D) distribution and volume-weighted projections showing heating rate estimates from these are displayed in Fig.~\ref{fig:heating_K_S}. The left panels show a 2D histogram of heating rates, where the color represents the volume fraction per logarithmic bin widths. The quadrant outlined by red dashed lines indicates where both approaches yielded positive values, indicating genuine (and physically correct) addition of heat. The distribution in both figures indicates that a large volume for both entropy-based approaches produces similar results in both runs. However, a non-negligible fraction of cells also shows discrepancies -- where $\mathcal{H}(s)<0$ and $\mathcal{H}(K)>0$, along with negative heating in several regions. Negative heating rates from both scalars also appeared in the test problems, for example, in the CR heating test (Sect.~\ref{sec:cr_heating}), but only in regions with no true heating (i.e., where the CR pressure gradient vanished). This suggests that the negative values from both approaches can arise from reconstruction errors in zones with no actual heat addition.

In the ``diffusive configuration'' (light jet with first-order reconstruction), both the discrepancy regions -- where $\mathcal{H}(s) < 0$ and $\mathcal{H}(K) > 0$ -- and the negative heating regions -- where $\mathcal{H}(s) < 0$ and $\mathcal{H}(K) < 0$ -- are significantly reduced. Remaining discrepancies occur primarily in regions where $\mathcal{H}(s) > 0$ but $\mathcal{H}(K) < 0$, typically associated with cells near contact discontinuities where advection errors are amplified, as discussed in Sect.~\ref{sec:post_contact}.

Thus, in our analysis, we consider the regions where both the $K$- and $s$-based methods yield positive heating, i.e., the areas enclosed by the red dashed lines. In Fig.~\ref{fig:heating_K_S}, within a spherical radius of 100~kpc, i.e., encompassing the jets, the total heating amounts to $2.9 \times 10^{45}~\ergs$ for the light case (left) and $2.3 \times 10^{45}~\ergs$ for the light: first-order (on the right), respectively. These values are in good agreement with the injected power in the domain, especially for the standard case.

The right panel in Fig.~\ref{fig:heating_K_S} displays the corresponding spatial distribution of heating rates for the light jet, comparing the standard run (left) with the ``diffusive configuration'' (right), using $\mathcal{H}(K)$. The maps show volume-weighted projections over a depth of 5~kpc, with cells exhibiting unphysical (negative) heating being omitted to highlight genuinely heated gas. When gradient extrapolations are disabled (right), the heating distribution becomes smoother and more diffuse, and the forward shock appears noticeably more uniform and broad. Discrepancies persist mainly along the jet--cocoon boundaries, i.e., at the contact discontinuities, where $\mathcal{H}(K)<0$ at several regions in both cases. In the ``diffusive configuration'' with only first-order reconstruction, no genuine heating is observed at or near these interfaces. In contrast, the standard run displays localized heating near these regions, including a mixture of strong and weak heating patches. 

In summary, the above discussion demonstrates that the discrepancies between the $K$- and $s$-based estimators are predominantly caused by advection errors and inaccuracies resulting from higher-order reconstruction in regions of the flow with strong gradients. In stochastically heated regions, such as around jets in clusters explored in our study, several numerical and discretization-induced discrepancies are often unavoidable. In such cases, the systematic approach outlined above enables us to filter out spurious heating contributions. Thus, it provides a robust method for identifying genuine physical heating despite numerical and flow-induced fluctuations.

\section{Turbulent heating estimation using velocity structure function and entropy-tracer based methods}
\label{app:heat_spectral_entro}

\subsection{Turbulent box simulations}
\label{app:vsf_turb}

\begin{figure}
\centering
\includegraphics[width=\linewidth, keepaspectratio]{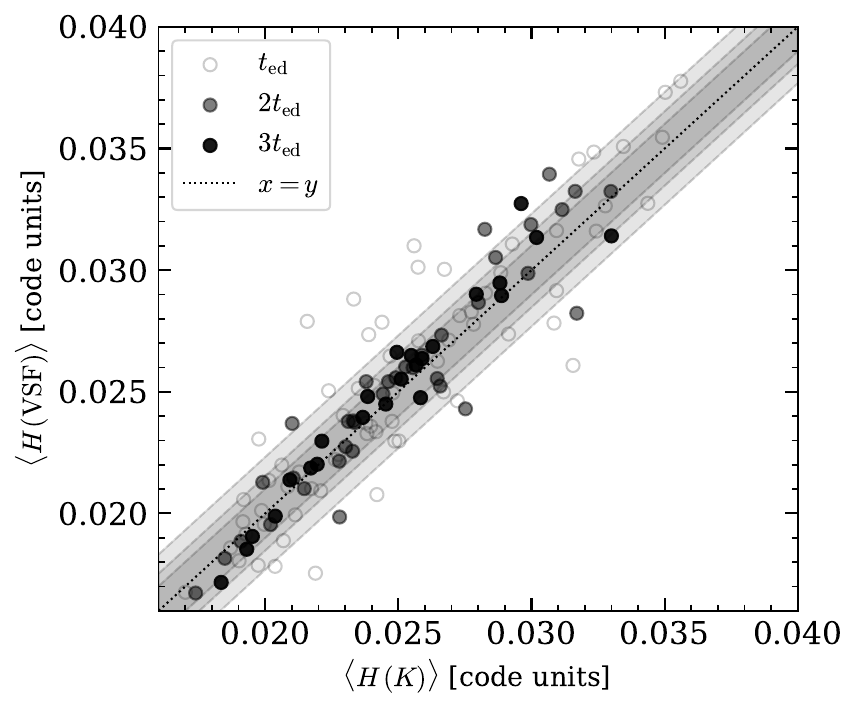}
\caption{Comparison of heating rates in a turbulent periodic box using the entropy tracer and VSF-based approach. The data points are averaged over different eddy turn-over times ($t_{\rm ed}$) to compare the convergence between the two methods. The shaded regions indicate the scatter (standard deviation) of each of the sets of data points.}
\label{fig:vsf_turb_box}
\end{figure}

In this section, we estimate the turbulent heating rate using our entropy-tracer-based heating estimator and compare it with the spectral approach employed in \citet{li_2025}. We benchmark the two methods using the idealized turbulent-box setup described in Sect.~\ref{sec:turb_heating}. In this setup, isotropic, homogeneous, subsonic turbulence is driven in a periodic domain on scales ranging from the box length ($L=1$ in code units) down to $L/2$. The simulation is evolved for 64 eddy turnover times (estimated at the injection scale) after reaching a statistically steady turbulent state. The forcing is random and statistically stationary, and therefore does not generate any net bulk flow. Under these conditions, the assumptions of Kolmogorov's theory for turbulence \citep{kolmogorov_1941} are expected to hold, with kinetic energy cascading from large to small scales and producing the characteristic inertial-range scaling. To estimate the turbulent dissipation rate following \citet{li_2025}, we compute the longitudinal second-order velocity structure function (VSF),
\begin{equation} 
D_{\ell \ell} (\ell)=\left\langle\left[\big(\bm{\varv}(\bm{x}+\bm{\ell})
-\bm{\varv}(\bm{x})\big)\bm\cdot\hat{\bm{\ell}}\right]^2\right\rangle,
\end{equation}
where $\bm{\varv}$ denotes the fluid velocities of two points, which are separated by the vector $\bm\ell$ of length $\ell$. Here $\bm{\hat{\ell}} = \bm{\ell}/\ell$ is the separation unit vector.
The computed VSFs exhibit a well-defined inertial range over $\ell \simeq 0.15 - 0.3$, where they are well described by the Kolmogorov scaling \citep{sreenivasan_1995}, \begin{equation} 
D_{\ell\ell} \simeq 2 (\varepsilon_\rmn{t} \ell)^{2/3}, 
\end{equation}
with $\varepsilon_\rmn{t}$ representing the characteristic turbulent energy dissipation rate per unit mass ($\ergs {\rm g}^{-1}$) inferred from the normalization of the fit. The corresponding volumetric heating rate was obtained by multiplying $\varepsilon_\rmn{t}$ by the mean density and integrating over the entire box volume. We estimate the heating rate after the quasi-steady turbulent state was established. In Fig.~\ref{fig:vsf_turb_box}, the time-averaged turbulent heating rates estimated using the VSF method and the entropy-tracer approach, over the whole domain, are compared. The agreement between the two methods improves substantially when the heating rates are averaged over more than one eddy turnover time. The standard deviation ($\sigma_{t_{\rm ed}}$) of the differences ($x-y$), as indicated using the filled regions ($\pm 1 \sigma_{t_{\rm ed}}$) around the $x=y$ line, decreases from $2.3\times 10^{-3}$ at $t_{\rm ed}$ to $9.6\times 10^{-4}$ at $3 t_{\rm ed}$. Also, the mean ratio remains close to unity for all values of $t_{\rm ed}$. This analysis implies that when employing spectral methods to infer turbulent dissipation rates, the heating should be averaged over more than one characteristic eddy turnover time to obtain a representative measure of the turbulent energetics.

\begin{figure*}
\centering
\includegraphics[width=0.43\linewidth, keepaspectratio]{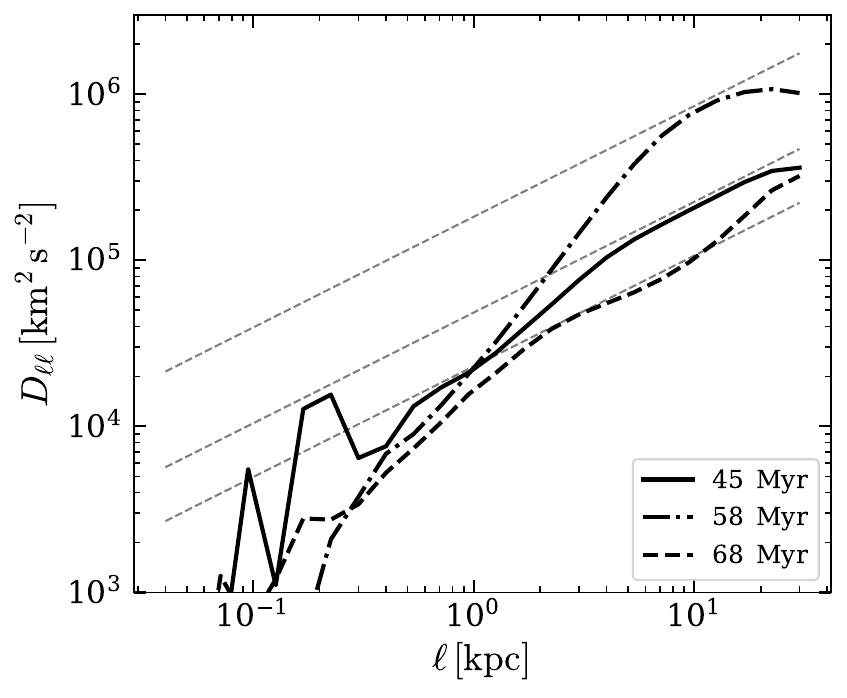}
\includegraphics[width=0.43\linewidth, keepaspectratio]{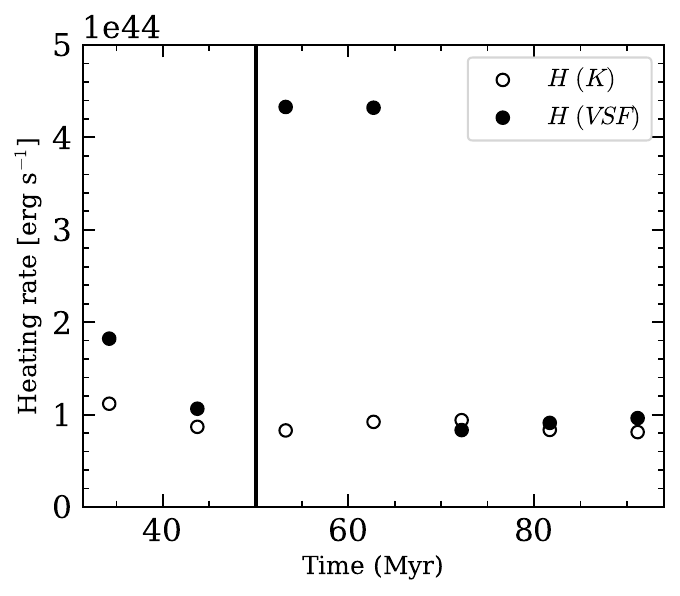}
\caption{\textit{Left}: Longitudinal velocity structure function (VSF) as a function of separation distance ($\ell$) computed for a 30~kpc spherical central region for the light jet case at different times. The dashed gray line is a reference for the Kolmogorov behaviour ($D_{\ell \ell} \propto \ell^{2/3}$) and is used to identify the inertial region in the VSFs. The jet is shut off at 50~Myr. \textit{Right}: Heating rate within the 30~kpc region, estimated using entropy-tracer-based and VSF-based approaches as a function of time (Myr). The values are averaged over a period of 10 Myr. The vertical line denotes the jet shutoff time.}
\label{fig:vsf_jet}
\end{figure*}

\subsection{AGN jet simulations}
We apply the VSF-based approach discussed above to the light jet simulation in this work, and compare the heating estimates with the entropy-based approach. The VSF is computed within the core region (radius < 30 kpc) on one side of the domain, selecting only gas with $T > 10^6$ K to exclude cold gas motions, and additionally restricting to regions with jet scalar $X_\rmn{jet}< 0.1$\footnote{Unlike the analysis in the main manuscript, where we adopt $X_{\rm jet}\leq0.9$ to estimate the heating outside the jet plasma, here we use a more restrictive threshold ($X_{\rm jet}\leq0.1$) to further exclude regions contaminated by the jet. This is because the jet plasma is dominated by highly anisotropic, coherent flows, for which the assumptions underlying Kolmogorov turbulence are not valid. Including these regions leads to large VSFs, particularly while the jet is active. The VSF from both cutoffs starts to converge only about 10 Myr after the jet is switched off.} to minimize contamination from the direct jet plasma. The jet is shut down at 50~Myr. The characteristic peak heating rate per unit mass is estimated from the portion of the inertial range that most closely follows the Kolmogorov scaling, assuming purely hydrodynamic turbulence for simplicity. The VSFs at three representative times are shown in the left panel of Fig.~\ref{fig:vsf_jet}, corresponding to one snapshot where the jet is active (45~Myr) and two snapshots after it switched off. For reference, the Kolmogorov scaling is overplotted, which is used to determine the maximum value of the inertial range for $\ell$. Shortly after the jet turns off, the VSF rises, reflecting the local bulk motions driven by the jet. At later times, the VSF instead shows an intermediate flattening followed by a rise at larger $\ell$. This seems to be evidence for two distinct populations of motion superposed within a single VSF: a local turbulent cascade at small $\ell$, and residual large-scale bulk motion at larger $\ell$. The panel on the right-hand side of Fig.~\ref{fig:vsf_jet} shows the total heating rate estimated from the two approaches, each averaged over a 10~Myr window.

Our analysis shows that the heating rates inferred from the two approaches in a jet-driven system can either agree or differ, depending on the dynamical state of the flow. In particular, the two methods closely agree around $\sim$45~Myr, by which time the jet has propagated over a height of $\sim$70~kpc, and at 75 Myr (i.e., approximately 25~Myr after shutdown). In contrast, significant discrepancies arise at other times, especially immediately following jet turn-off. This indicates that coherent motions in the central region, which become particularly prominent just after the jet switched off, can substantially enhance the VSF from this region without necessarily corresponding to fully developed turbulence (e.g., see the $t=58~$Myr curve in the left panel). This is consistent with the suggestion of \citet{li_2025} that the temporary increase in VSF amplitude may reflect bulk motions rather than fully developed turbulence, and that heating rates estimated from the VSF -- which can carry a contribution from these bulk motions -- may consequently overestimate the true turbulent heating rate. In line with this, the VSF-based method yields heating rates significantly higher than those from the entropy-tracer approach during this phase. As the gas motions are randomized with time, the VSF-derived heating rate decreases and converges toward the entropy-based estimate, becoming consistent after roughly 75~Myr (i.e., 25 Myr after the jet is switched off). Thus, the benchmark presented in \ref{app:vsf_turb}, where both approaches agree well, represents an idealized case of homogeneous, isotropic, and steady turbulence with a well-defined inertial range. These assumptions generally do not hold in more realistic astrophysical environments, such as AGN jet simulations in galaxy clusters, where the flow is highly inhomogeneous, anisotropic, and continuously influenced by large-scale bulk motions. Consequently, applying spectral methods as above to such systems is considerably more challenging. In the future, we plan to explore filtering-based approaches as developed in \citet{perrone_2026} to deduce the turbulent dissipation in our simulations.

\section{Helmholtz decomposition in AGN jet simulations}
\label{sec:helm_decompose}
\begin{figure*}
\centering
\includegraphics[width=0.95\linewidth, keepaspectratio]{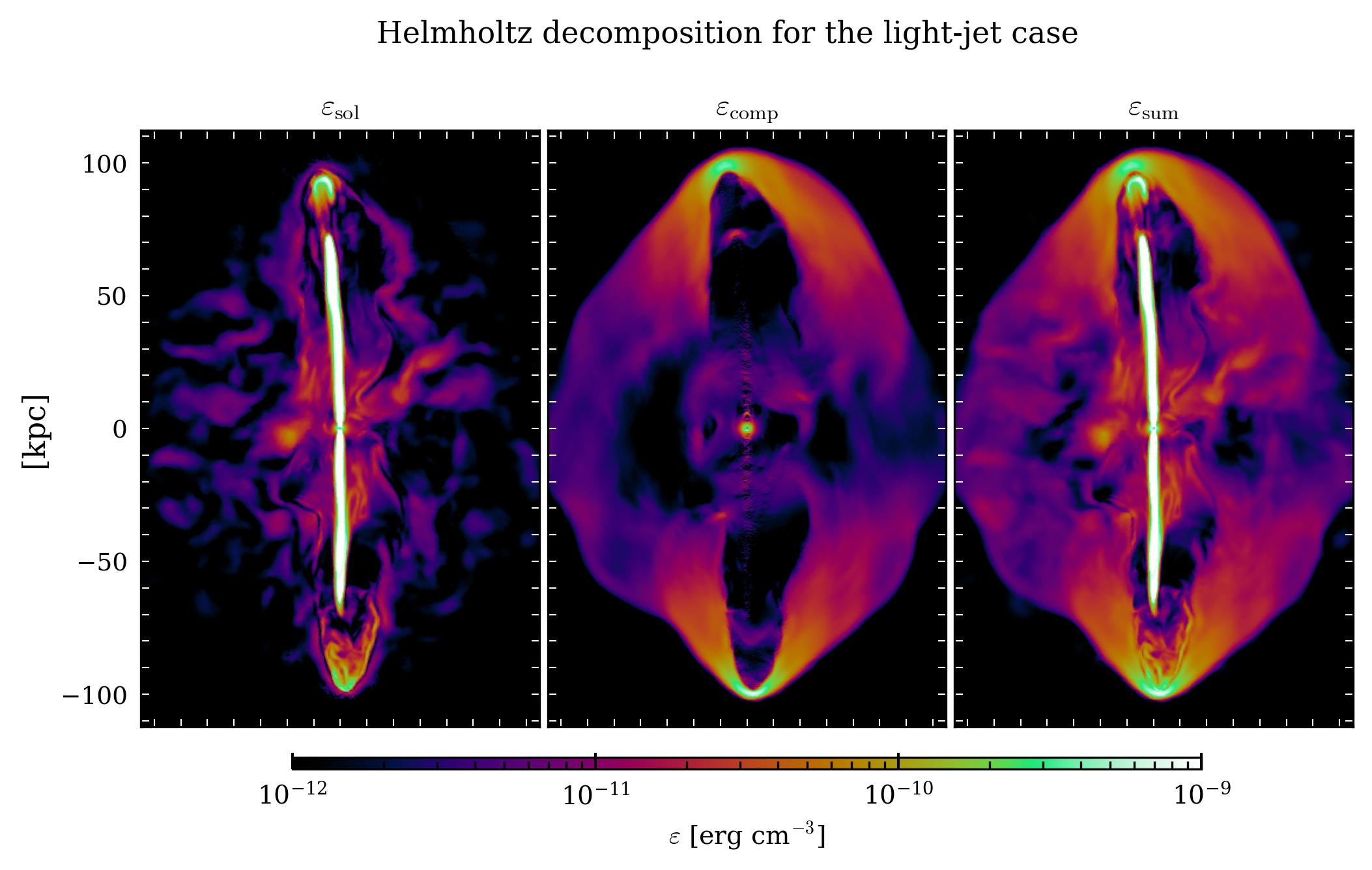}
\caption{Helmholtz decomposition separating the kinetic energy density into different modes for the light-jet case at $t = 49~\mathrm{Myr}$, i.e., the snapshot before the jet was switched off. Shown are the kinetic energy densities for the solenoidal (\textit{left}), compressible (\textit{middle}) components, and the sum of both components (\textit{right}).}
\label{fig:decompose_ldj}
\end{figure*}

We perform Helmholtz decomposition to separate the compressible and solenoidal components of the turbulent flow in the AGN-jet simulations. The decomposition is carried out on a uniform Cartesian grid covering a cubic domain of side length 225 kpc, which encompasses the maximum extent of the jet during its active phase. The Voronoi-based \textsc{Arepo} fields, including the velocity components, gas density, heating-rate density, and jet tracer, are interpolated onto a $512^3$ uniform grid, corresponding to a spatial resolution of 440 pc. Although this is coarser than the resolution of the jet region (300 pc) in our simulations, we verify that the interpolation introduces negligible information loss; i.e., the total mass, kinetic energy, and heating rate within the analysis volume agree between the Voronoi and uniform grids to within a few per cent.

The interpolation is performed using the 10 nearest Voronoi cells and a
distance-dependent kernel,
\begin{equation}
W(\bm{r}_i - \bm{r}_j,h)=1-\left(\frac{|\bm{r}_i-\bm{r}_j|}{h}\right)^2, \qquad |\bm{r}_i-\bm{r}_j|\leq h,
\end{equation}
where $h=1.5 r_{\rm max}$ and $r_{\rm max}$ is the distance to the tenth
nearest neighbour. For each quantity on the Voronoi mesh ($f_j$), the value at a uniform-grid point ($f_{{\rm grid},i}$) is computed with an SPH-like weighted average,
\begin{equation}
f_{{\rm grid},i} =
\frac{\displaystyle\sum_j W (\bm{r}_i-\bm{r}_j,h) f_j}
{\displaystyle\sum_j W (\bm{r}_i-\bm{r}_j,h)}.
\end{equation}

Thereafter, we perform a fast Fourier transform (FFT) of the density-weighted velocity field, particularly $ \bm{\varw} (\bm{x})=\sqrt{\rho(\bm{x})} \bm\varv(\bm{x})$ \citep{kritsuk_2007}, where

\begin{align}
    \bm{\tilde{\varw}}(\bm{k}) = \frac{1}{(2\pi)^3} \int_V \bm{\varw}(\bm{x}) \rmn{e}^{-2\pi {\rm i} \bm{k \cdot x}} {\rm d}^3{x}
\end{align}
gives the transformed field in Fourier space.
Using Helmholtz decomposition, we separate the compressible and solenoidal kinetic modes as
\begin{align}
\bm{\tilde{\varw}}_{\rm comp}(\bm{k})
    &= \frac{\bm{k}\,\left(\bm{k}\cdot\bm{\tilde{\varw}}(\bm{k})\right)}
            {|\bm{k}|^2},
    \quad \text{and} \\
\bm{\tilde{\varw}}_{\rm sol}(\bm{k})
    &= \bm{\tilde{\varw}}(\bm{k}) - \bm{\tilde{\varw}}_{\rm comp}(\bm{k}) .
\end{align}
The inverse Fourier transform is then used to recover the density-weighted velocity components in real space, which implies a compressible (i.e., curl-free: $\bnabla \bm{\times} \bm{\varw}_{\rm comp} = \textbf{0}$) and a solenoidal component (i.e., divergence-free: $\bnabla \bcdot \bm{\varw}_{\rm sol} = 0$). Since the FFT implicitly assumes periodic boundary conditions, this assumption may not be strictly valid for our domain here, particularly after the jets are switched off, and the resulting outflows propagate to distances comparable to or larger than the selected region. To minimise potential boundary effects, we restrict our heating analysis to the inner 200 kpc region while adopting a larger, 225 kpc cubic domain for the Fourier decomposition. Thus, the boundaries of the FT domain lie outside the region used for the heating analysis.

In a periodic domain, the kinetic energy equals the sum of solenoidal and compressible energies,\footnote{Note that the cross term of density-weighted compressible and solenoidal velocities vanishes identically for a periodic domain, since $\int_V \bm{\varw}_{\rm sol} \bm\cdot\bm{\varw}_{\rm comp}\, {\rm d}V = \int_V \bnabla \bm\cdot(\Phi\bm{\varw}_{\rm sol})\, {\rm d}V = \int_{\partial V} \Phi\bm{\varw}_{\rm sol} \bm\cdot\rmn{d}\bm{A} =0$, where $\bm{\varw}_{\rm comp}=\bnabla\Phi$ and because the surface flux of the periodic vector field through the opposing faces of the cubic domain cancel exactly.}
\begin{align}
E_\rmn{kin}&=\int_V \frac{1}{2} \bm{\varw}^2 {\rm d}V = \frac{1}{2}\int_V \left(\bm{\varw}_{\rm sol} + \bm{\varw}_{\rm comp} \right)^2 {\rm d}V \nonumber\\
&= \int_V (\varepsilon_{\rm sol} + \varepsilon_{\rm comp}) \, {\rm d}V = E_\rmn{sol}+E_\rmn{comp}.
\end{align}
As a consistency check, we verify that this relation holds in real space with a relative difference of less than $0.1\%$ when integrated over the analysis domain, computed at all times. This demonstrates that the Helmholtz decomposition and subsequent inverse Fourier transform recover the kinetic-energy content of the original velocity field with negligible numerical error, and that any numerical cross-terms introduced by the decomposition do not make a significant contribution. In Fig.~\ref{fig:decompose_ldj}, we show the kinetic energy densities of the individual components ($\varepsilon_{\rm sol}$ and $\varepsilon_{\rm comp}$) in the $x-y$ plane for the light-jet case at 49~Myr, and the sum of both components ($\varepsilon_{\rm sum}$). This corresponds to the snapshot whose maps of density, jet-scalar, pressure, and heating rate density are shown in the middle row in Fig.~\ref{fig:heat_collage_1}. Our analysis indicates that the solenoidal components ($\varepsilon_{\rm sol}$) dominate within the lobes (i.e., away from the jet's spine), whereas compressible motions ($\varepsilon_{\rm comp}$) are stronger within the shocked ambient medium. Also, the shocked ambient regions show localized enhancement in the solenoidal component, which indicates that the shocked plasma can also develop vortical motions after it crosses the curved forward shock according to Crocco's theorem \citeyearpar{Crocco}. In our study (see Sect.~\ref{sec:jet_test}), we use the fraction of the kinetic energies in the two kinetic modes to deduce their contribution to the localized heating rates.

\label{LastPage}

\end{document}